\documentclass{aa}
\usepackage[usenames,dvipsnames]{xcolor}
\usepackage{graphicx}
\graphicspath{{./figs/}}
\usepackage{txfonts}
\usepackage{gensymb}
\usepackage{tikz}
\usepackage{courier}
\usepackage{afterpage}
\usepackage{newtxtext}
\usepackage[frenchmath,varg]{newtxmath}
\usepackage{natbib,twoopt}
\usepackage[breaklinks=true,colorlinks=true,linkcolor=NavyBlue!50!Emerald,citecolor=NavyBlue!50!Emerald,filecolor=blue,urlcolor=NavyBlue!50!Emerald]{hyperref}
\usepackage[para,online,flushleft]{threeparttable}

\newcommand*\rfrac[2]{{}^{#1}\!/_{#2}}%running fraction with slash - requires math mode.

\bibpunct{(}{)}{;}{a}{}{,} %% natbib format for A&A and ApJ
\makeatletter

\makeatletter
\newcommandtwoopt{\citeads}[3][][]{\href{http://adsabs.harvard.edu/abs/#3}%
{\def\hyper@linkstart##1##2{}%
\let\hyper@linkend\@empty\citealp[#1][#2]{#3}}}
\newcommandtwoopt{\citepads}[3][][]{\href{http://adsabs.harvard.edu/abs/#3}%
{\def\hyper@linkstart##1##2{}%
\let\hyper@linkend\@empty\citep[#1][#2]{#3}}}
\newcommandtwoopt{\citetads}[3][][]{\href{http://adsabs.harvard.edu/abs/#3}%
{\def\hyper@linkstart##1##2{}%
\let\hyper@linkend\@empty\citet[#1][#2]{#3}}}
\newcommandtwoopt{\citeyearads}[3][][]%
{\href{http://adsabs.harvard.edu/abs/#3}
{\def\hyper@linkstart##1##2{}%
\let\hyper@linkend\@empty\citeyear[#1][#2]{#3}}}
\makeatother

\def\micron{\hbox{$\mu$m}}

\begin{document} 

   \titlerunning{Compact jets in LLAGN}
   \authorrunning{Fern\'andez-Ontiveros, L\'opez-L\'opez \& Prieto}
   \title{Compact jets dominate the continuum emission in low-luminosity active galactic nuclei}

   % \subtitle{}
   % \thanks{Based on VLT XX.X-XXXA, YY.Y-YYYYA}

   \author{J.\,A. Fern\'andez-Ontiveros\inst{1,2}\thanks{\email{\sf \href{mailto:j.a.fernandez.ontiveros@gmail.com}{j.a.fernandez.ontiveros@gmail.com}, \href{mailto:jafernandez@cefca.es}{jafernandez@cefca.es}}}, %\fnmsep
          X. L\'opez-L\'opez\inst{3,4,5},
          A. Prieto\inst{6,3}
          }

   \institute{Centro de Estudios de F\'isica del Cosmos de Arag\'on (CEFCA), Plaza San Juan 1, E--44001 Teruel, Spain
%\\\email{\sf \href{mailto:j.a.fernandez.ontiveros@gmail.com}{\color{JungleGreen!70!Black}j.a.fernandez.ontiveros@gmail.com}, \href{mailto:juan.fernandez@inaf.it}{\color{JungleGreen!70!Black}juan.fernandez@inaf.it}}
    \and
    Istituto di Astrofisica e Planetologia Spaziali (INAF--IAPS), Via Fosso del Cavaliere 100, I--00133 Roma, Italy 
    \and
    Universidad de La Laguna (ULL), Dpto. Astrof\'isica, Avd. Astrof\'isico Fco. S\'anchez s/n, E--38206 La Laguna, Tenerife, Spain
    \and
    Dipartimento di Fisica e Astronomia, Universit\`a degli Studi di Bologna, via Gobetti 93/2, I--40129 Bologna, Italy
    \and
    INAF--OAS, Osservatorio di Astrofisica e Scienza dello Spazio di Bologna, via Gobetti 93/3, I--40129 Bologna, Italy
    \and
    Instituto de Astrof\'isica de Canarias (IAC), C/V\'ia L\'actea s/n, E--38205 La Laguna, Tenerife, Spain
   }

   \date{Received March 14, 2022; accepted November 11, 2022} %\today

% \abstract{}{}{}{}{} 
% 5 {} token are mandatory

\abstract{Low-luminosity active galactic nuclei (LLAGN) are special among their kind due to the profound structural changes that the central engine experiences at low accretion rates ($\lesssim 10^{-3}$ in Eddington units). The disappearance of the accretion disc --\,the blue bump\,-- leaves behind a faint optical nuclear continuum whose nature has been largely debated. This is mainly due to serious limitations on the observational side imposed by the starlight contamination from the host galaxy and the absorption by hydrogen, preventing the detection of these weak nuclei in the infrared (IR) to ultraviolet (UV) range. We addressed these challenges by combining multi-wavelength sub-arcsecond resolution observations --\,able to isolate the genuine nuclear continuum\,-- with nebular lines in the mid-IR, which allowed us to indirectly probe the shape of the extreme UV continuum. We found that eight of the nearest prototype LLAGN are compatible with pure compact jet emission over more than ten orders of magnitude in frequency. This consists of self-absorbed synchrotron emission from radio to the UV plus the associated synchrotron self-Compton component dominating the emission in the UV to X-ray range. Additionally, the LLAGN continua show two particular characteristics when compared with the typical jet spectrum seen in radio galaxies: \textit{i)} a very steep spectral slope in the IR-to-optical/UV range ($-3.7 < \alpha_0 < -1.3$; $F_\nu \propto \nu^{\alpha_0}$); and \textit{ii)} a very high turnover frequency ($0.2$--$30\, \rm{THz}$; $1.3\, \rm{mm}$--$10\, \rm{\micron}$) that separates the optically thick radio emission from the optically thin continuum in the IR-to-optical/UV range. These attributes can be explained if the synchrotron continuum is mainly dominated by thermalised particles at the jet base or the corona with considerably high temperatures, whereas only a small fraction of the energy ($\sim 20\%$) would be distributed along the high-energy power-law tail of accelerated particles. On the other hand, the nebular gas excitation in LLAGN is in agreement with photo-ionisation from inverse Compton radiation ($\alpha_{\rm x} \sim -0.7$), which would dominate the nuclear continuum shortwards of $\sim 3000$\,\AA, albeit a possible contribution from low-velocity shocks ($< 500\, \rm{km\,s^{-1}}$) to the line excitation cannot be discarded. No sign of a standard hot accretion disc is seen in our sample of LLAGN, nevertheless, a weak cold disc ($< 3000\, \rm{K}$) is detected at the nucleus of the Sombrero galaxy, though its contribution to the nebular gas excitation is negligible. Our results suggest that the continuum emission in LLAGN is dominated at all wavelengths by undeveloped jets, powered by a thermalised particle distribution with high energies, on average. This is in agreement with their compact morphology and their high turnover frequencies. This behaviour is similar to that observed in peaked-spectrum radio sources and also compact jets in quiescent black hole X-ray binaries. Nevertheless, the presence of extended jet emission at kiloparsec scales for some of the objects in the sample is indicative of past jet activity, suggesting that these nuclei may undergo a rejuvenation event after a more active phase that produced their extended jets. These results imply that the dominant channel for energy release in LLAGN is mainly kinetic via the jet, rather than the radiative one. This has important implications in the context of galaxy evolution, since LLAGN probably represent a major but underestimated source of kinetic feedback in galaxies.}


\maketitle

\section{Introduction}\label{intro}

The majority of active galactic nuclei (AGN) exhibit a low level of activity characterised by modest bolometric luminosities ($\lesssim 10^{43}\,\rm{erg\,s^{-1}}$) and low accretion rates ($\sim 10^{-3}\, \rm{\dot{M}_{\rm Edd}}$), in contrast with the luminous Seyfert galaxies and quasars ($\gtrsim 10^{43}\,\rm{erg\,s^{-1}}$; $\sim 0.01$--$0.1\, \rm{\dot{M}_{\rm Edd}}$). Known as low-luminosity AGN (LLAGN), these faint nuclei are found in approximately one-third of all galaxies in the local Universe \citep{ho2008}, albeit this fraction is largely dependent on the selection criteria used to identify LLAGN \citep{goulding2009,capetti2011,bar2017}. However, LLAGN are not mere scale-down versions of their bright counterparts, showing significant differences in the nature of the nuclear continuum and the properties of the galaxies that host them: \textit{i)} they miss the big blue bump \citep{ho1996b}, the spectral signature of the thermal emission from the accretion disc in the optical/UV, and the IR bump \citep[e.g.][this work]{prieto2016,reb2018}, characteristic of reprocessed disc emission by dust; \textit{ii)} they are frequently associated with low-ionisation nuclear emission-line regions (LINERs, \citealt{heckman1980}); \textit{iii)} most LLAGN are radio-loud, showing compact cores and parsec-scale radio jets \citep{nagar2005,mezcua2014,baldi2018,baldi2021}; and \textit{iv)} the majority are hosted by elliptical and early-type spiral galaxies \citep{ho1996a,ho2008}.

The main differences in the nuclear continuum between bright Seyferts and LLAGN are likely caused by profound structural changes in the central engine happening at low accretion rates. For instance, the broad-line region would no longer be supported by radiation pressure from the nucleus below $10^{-3}\, \rm{L_{Edd}}$ \citep{nicastro2003,elitzur2009}, while the dusty torus is also expected to vanish in this regime \citep{elitzur2006}. The absence of the blue bump emission led to the use of advection-dominated accretion flow (ADAF) models \citep{narayan1994}, which incorporate a truncated accretion disc in their inner radius to explain the low radiative efficiency in supermassive black holes (BHs) accreting at very low Eddington rates (e.g. Sgr\,A$^*$ in \citealt{narayan1995}, \citealt{fabian1995}; LLAGN in \citealt{2009ApJ...703.1034Y}, \citealt{eracleous2010}, \citealt{yuan2014}). Additionally, the continuum emission in these nuclei seems to be strongly determined by the coupling between accretion and ejection in the jet-corona plus disc system \citep{blandford1999}. This is further supported by the fundamental plane of BH accretion, where LLAGN lie together with jet-dominated AGN and X-ray binaries \citep{merloni2003,falcke2004,koerding2006a}. 
%, a tight relation between the radio and X-ray luminosities followed by accreting BHs across the mass spectrum when the energy output is dominated by the jet-corona. %, once the proper mass normalisation is taken into account.

\begin{figure}
\centering
\includegraphics[width = \columnwidth]{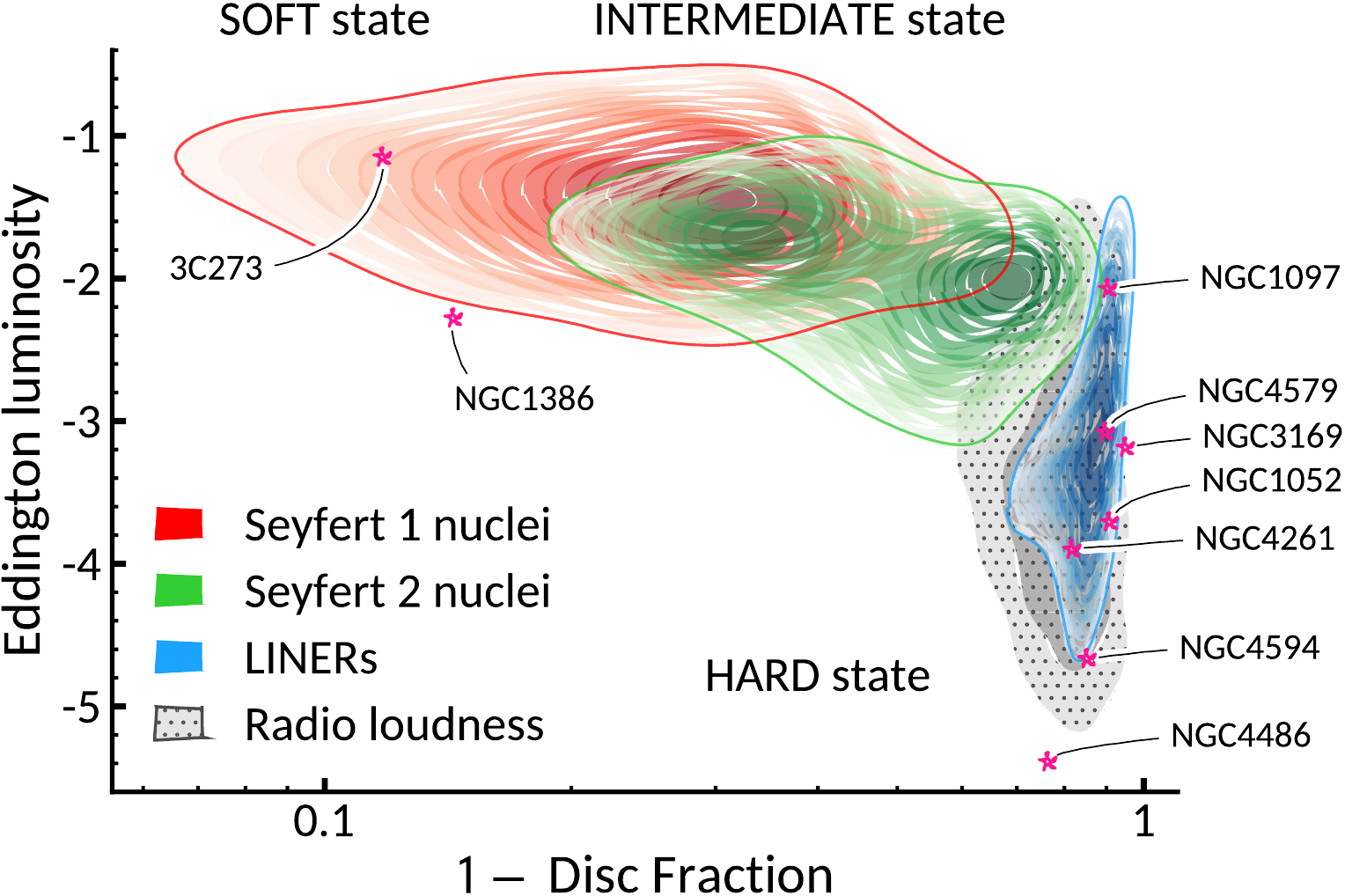}
\caption{Luminosity-Excitation Diagram (LED) from \citet{jafo2021} indicating the location of the sources analysed in this study. Accretion-disc dominated sources in the soft state (e.g. 3C\,273) are found in the upper-left side of the LED, while corona/jet-dominated sources in the hard state are found on the right side of the LED. The presence of strong shocks in the case of NGC\,1386 \citep{rodriguez-ardila2017} significantly enhance the flux of the [\ion{O}{iv}]$_{25.9}$ line, causing a shift of this source to the left side of the LED. We note that the Eddington luminosities in this figure are based only on the luminosity of the [\ion{O}{iv}]$_{25.9}$ and [\ion{Ne}{ii}]$_{12.8}$ lines, for consistency with the LED diagram in \citet{jafo2021}.}\label{fig_led}
\end{figure}

Bright radio emission dominated by a compact jet is a common characteristic of stellar-mass BHs at low Eddington rates ($\lambda_{\rm Edd} = \log(L_{\rm bol}/L_{\rm Edd}) \lesssim -3$), that is in the hard state \citep{fender2001}, while several evidences for low-accreting supermassive BHs in LLAGN point in the same direction \citep[e.g.][]{koerding2006b,markoff2008,prieto2016,tomar2021}. For a system in the hard state, the low accretion rate can no longer sustain the luminosity of the disc, which cools significantly \citep[e.g.][]{plant2015}. Subsequently, the corona increases its temperature due to the dearth of UV seed photons available for cooling through Inverse Compton processes, boosting the jet emission that overpowers the cold disc component \citep[e.g.][]{kalemci2013,cheng2020}. Contrarily, the bright thermal emission from the disc during the soft state efficiently cools the corona, turning the jet off \citep{dunn2010,meyer2012}. The existence of these accretion states is well known in stellar-mass BHs thanks to the hardness-intensity \citep{homan2001} or the hardness-luminosity diagrams \citep[e.g.][]{dunn2010}, while a similar phenomenology is observed in LLAGN, for example, shown by the disc fraction-luminosity diagram \citep{koerding2006b} and the luminosity-excitation diagram \citep[LED;][]{jafo2021}. The latter is based on the differential excitation of the infrared (IR) nebular transitions in the narrow-line gas when the disc contribution is prominent. LLAGN are found at the tail of the q-shaped diagram in the LED (Fig.\,\ref{fig_led}), corresponding to systems with low $\rm{\lambda_{Edd}}$ and low disc-fraction values, as expected for accreting supermassive BHs in the hard state \citep[see][for further details on the LED]{jafo2021}.

The nature of the continuum emission in LLAGN is still unclear and continues to be widely debated, even after a vast observational and modelling effort \citep{ho1999,ulvestad2001,laor2003,eracleous2010,pian2010,mason2012,prieto2016,jafo2019}. LLAGN continua have been modelled using ADAFs and jets with ambiguous results \citep{yu2011,nemmen2014}, largely due to the sparse sampling of the spectral energy distributions (SEDs) used as input for the fits, especially in the IR to UV range. This part of the electromagnetic spectrum is seriously affected by the starlight contamination from the host galaxy that outshines the nuclear emission by typically more than an order of magnitude. Therefore, the use of high-angular resolution techniques is required in this spectral range even for moderately luminous Seyfert galaxies \citep[e.g.][]{prieto2010}, and becomes mandatory for LLAGN. Furthermore, hydrogen atoms in the interstellar medium (ISM) of galaxies absorb most of the photons with energies above the Lyman break at $13.6\, \rm{eV}$ ($912$\,\AA) up to $\sim 0.1\, \rm{keV}$, extinguishing completely the nuclear continuum emission in this energy range.

The lack of insight on the nature of the continuum radiation in LLAGN led to a wide variety of ionising mechanisms that have been proposed in the literature to explain the nebular properties of these sources. These include massive stars \citep{terlevich1985,shields1992,colina2002}, shocks \citep{heckman1980,dopita1995,sugai2000}, and a hard nuclear continuum source alike to Seyfert nuclei with weaker radiation fields \citep{ferland1983,keel1983a,gabel2000}, which could be powered by a thin accretion disc surviving at low $\rm{\lambda_{Edd}}$ \citep{maoz2007}. A considerable contribution to the confusion comes from the fact that most LLAGN belong to the LINER spectral class, which defines a rather heterogeneous group including also non-AGN sources and off-centred emission regions in Seyfert galaxies. In these cases the nebular excitation can be explained by post-asymptotic giant branch stars and by the combination of different ionisation sources such as starbursts and AGN, respectively \citep[e.g.][]{yan2012,belfiore2016}.

\begin{table*}
\scriptsize
\centering
\setlength{\tabcolsep}{3.pt}
\caption{Sample of galaxies sorted by decreasing Eddington rate $\lambda_{\rm Edd}$. The spatial resolution is estimated as the \textsc{fwhm} of the most compact object found in the FOV. In all cases, the Galactic extinction has been corrected using the reddening values from \citet{schlafly2011} and the extinction curve from \citet{cardelli1989}. L$_{\rm X}$ is the luminosity in the $2$--$10\, \rm{keV}$ range, scaled to our adopted distance and corrected from absorption. L$_{\rm bol}$ is the bolometric luminosity extracted from the high-resolution continuum flux density distributions. L$_{\rm bol}^{\rm mod}$ is the integrated luminosity derived from the best-fit broken power law. In the case of NGC\,4594, L$_{\rm bol}^{\rm mod}$ includes also a thermal component to account for the cold disc emission peaking at $1\, \rm{\micron}$.}\label{tab_sample}
\resizebox{\textwidth}{!}{
\begin{tabular}{lllcccccccccccc}
\bf Name &
\bf Type &
\bf Host &
\bf D &
\bf RefD &
\bf log(M$_{\rm BH}$) &
\bf RefBH &
\bf log(L$_{\rm X}$) &
\bf RefX &
\bf log(L$_{\rm bol}$) &
\bf log(L$_{\rm bol}^{\rm mod}$) &
\bf $\lambda_{\rm Edd}^{\rm mod}$ &
\bf FWHM &
\bf 1$''$ &
\bf $A_{\rm V}$ \cr
 & & & (Mpc) & & ($\rm{M_\odot}$) & & ($\rm{erg\,s^{-1}}$) &  & ($\rm{erg\,s^{-1}}$) & ($\rm{erg\,s^{-1}}$) & & (pc) & (pc) & (mag) \cr
\hline\\[-0.2cm]
NGC 1386            & Sy2   & SB(s)a     & 15.3 & JEN03 & $7.72$ & HLED  & $42.15$ & MAS16 & $42.34$  & $42.65$ & $-3.17$   & 6.7  & 74.2  & 0.034 \cr
NGC 1097            & LINb  & SB(r$'$l)b & 14.2 & TUL09 & $7.17$ & DAV07 & $40.63$ & NEM06 & $41.79$  & $41.89$ & $-3.38$   & 8.9  & 68.8  & 0.073 \cr
NGC 1052            & LINb  & E4         & 18.0 & JEN03 & $8.56$ & HLED  & $41.33$ & GON09 & $42.84$  & $42.88$ & $-3.78$   & 10.5 & 87.3  & 0.073 \cr
NGC 4579 (M58)      & LINb  & SAB(rs)b   & 16.7 & TUL13 & $8.13$ & HLED  & $41.39$ & GON09 & $42.41$  & $42.35$ & $-3.88$   & 6.3  & 81.0  & 0.112 \cr
NGC 3169            & LIN   & SA(s)a     & 24.7 & MAN09 & $8.38$ & HO09  & $41.61$ & TER03 & $42.10$  & $42.50$ & $-3.98$   & 10.9 & 119.7 & 0.085 \cr
NGC 4261 (3C\,270)    & LINb  & E2-3       & 32.4 & TUL13 & $8.72$ & FER96 & $41.22$ & GON09 & $42.24$  & $42.51$ & $-4.31$   & 12.3 & 157.1 & 0.049 \cr
M87 (3C\,274)         & LINb  & cD         & 16.7 & BLA09 & $9.79$ & GEB11 & $40.49$ & PRI16 & $42.45$  & $42.32$ & $-5.57$   & 9.2  & 81.0  & 0.063 \cr
NGC 4594 (Sombrero) & LIN   & SA(s)a     & 9.08 & JEN03 & $8.82$ & JAR11 & $40.00$ & GON06 & $41.65$  & $41.40$ & $-5.70$ & 3.1  & 44.0  & 0.140 \cr
NGC 404             & LIN   & SA(s)0     & 3.05 & DAL09 & $5.65$ & SET10 & $37.13$ & PAR14 & $<40.37$ & --      & $< -3.38$ & 1.2  & 14.8  & 0.160 \cr
\end{tabular}}
\tablefoot{References for the redshift-independent distances. BLA09: \citealt{blakeslee2009}; DAL09: \citealt{dalcanton2009}; JEN03: \citealt{jensen2003}; MAN09: \citealt{mandel2009}; TUL09: \citealt{tully2009}; TUL13: \citealt{tully2013}. References for the BH mass estimates. FER96: \citealt{ferrarese1996}; GEB11: \citealt{gebhardt2011}; HLED, DAV07, HO09: based on the velocity dispersion given by HyperLEDA \citep{makarov2014}, \citet{davies2007}, or by \citet{ho2009}, using the M$_{\rm BH}$--$\sigma$ relation from \citet{kormendy2013}; JAR11: \citealt{jardel2011}; SET10: \citealt{seth2010}. References for the X-ray luminosities. PAR14: \citealt{paragi2014}; GON06: \citealt{gonzalez-martin2006}; NEM06: \citealt{nemmen2006}; MAS16: \citealt{masini2016}; PRI16: \citealt{prieto2016}; GON09: \citealt{gonzalez-martin2009b}; TER03: \citealt{terashima2003}.}
\end{table*}

This work examines the nature of the continuum emission in a sample of nine of the nearest ($3$--$32\, \rm{Mpc}$), best prototype LLAGN, covering a wide range in Eddington rate ($-5.5 < \lambda_{\rm bol} < -3$; see Fig.\,\ref{fig_diag}). Due to the aforementioned limitations of previous studies in the IR-to-optical/UV, a critical aspect is to determine the shape of the nuclear continuum in this range. This has been addressed following a twofold strategy that combines direct observations of the nuclear continuum, using high-angular resolution data, with an indirect probe of the extreme UV (EUV) continuum shape, inferred from the nebular lines in the mid-IR range observed by the Infrared Spectrograph \citep[IRS][]{houck2004} onboard the \textit{Spitzer Space Telescope} \citep{werner2004}. The first dataset includes space-based imaging with the \textit{Hubble Space Telescope (HST)}, adaptive-optics observations acquired with NaCo at the Very Large Telescope (VLT), and diffraction-limited imaging in the mid-IR with VLT/VISIR. These provide a direct measurement of the faint nuclear continuum emission avoiding the starlight contamination from the host galaxy. The second approach allows us to recover, at first order, the shape of the ionising continuum in the EUV, where direct measurements are not possible due to heavy hydrogen absorption.

The high-angular resolution observations and the spectroscopic data are described in Section\,\ref{data}. In Section\,\ref{results} we present the LLAGN continuum measurements and the analysis of the nebular ionising continuum using photo-ionisation models. The results are discussed in Section\,\ref{discuss} and the final summary is presented in Section\,\ref{summary}.

\section{Dataset}\label{data}

This project is a follow-up of \textit{The central parsecs of the nearest galaxies}\footnote{\url{http://research.iac.es/project/parsec}} \citep{reunanen2010,prieto2010,jafo2011a}, a high-spatial resolution study of the brightest and nearest Seyfert galaxies carried out at sub-arcsecond scales with the VLT. The objects included in the present work are about one to two orders of magnitude fainter, they are 9 LLAGN selected among the nearest ones, thus suitable for adaptive optics (AO) observations (see Table\,\ref{tab_sample}). Their nuclei are not covered by dust lanes (except NGC\,1386 and NGC\,3169), thus the absence of the blue bump is genuine. Three of the objects are the canonical reference for definition of the LLAGN class (NGC\,1052 in \citealt{heckman1980}; NGC\,1097 in \citealt{keel1983b}; M87 in \citealt{fabian1995}). NGC\,1097 is the source with the highest luminosity in Eddington units and showed a transition to a Seyfert 1 detected in November 1991 till January 1994 \citep{storchi-bergmann1993,storchi-bergmann1995}. The sample also represents the low-luminosity class in terms of host galaxy (S0, E and Sa: e.g. M87, NGC\,1052 and NGC\,3169) and radio loudness/quietness (e.g. M87 versus NGC\,1097).

\begin{table*}
\caption{Fluxes of the mid-IR nebular lines for the galaxies analysed in this work (in $10^{-17}\, \rm{W\,m^{-2}} \equiv 10^{-14}\, \rm{erg\,cm^{-2}\,s^{-1}}$). All fluxes have been extracted from \textit{Spitzer}/IRS spectra acquired with the high-spectral resolution mode ($R = 600$) obtained from the CASSIS database and the \textit{Spitzer} Heritage Archive.}
\centering
\label{table_IR_lines}
% For LaTeX tables you can use
\begin{tabular}{lccccc}

\bf Object &
\bf [\ion{S}{iv}]$_{\rm 10.5 \mu m}$  &
\bf [\ion{Ne}{ii}]$_{\rm 12.8\mu m}$  &
\bf [\ion{Ne}{iii}]$_{\rm 15.6\mu m}$ &
\bf [\ion{S}{iii}]$_{\rm 18.7\mu m}$  &
\bf [\ion{O}{iv}]$_{\rm 25.9\mu m}$   \cr
\hline\\[-0.2cm]
%\mr
NGC\,1386    & $24.3 \pm 0.8$& $11.8 \pm 0.4$ & $38.1 \pm 0.6 $ & $16.2\pm 0.6$ & $97   \pm 5$\tablefootmark{a}   \cr
NGC\,1097    & $ <0.45 $     & $12   \pm 2  $ & $1.7  \pm 0.5 $ & $8.4 \pm 0.5$ & $3    \pm 1$   \cr
NGC\,1052    & $ <1.2 $      & $22.4 \pm 0.6$ & $12.3 \pm 0.7 $ & $4.8 \pm 0.8$ & $2.4  \pm 0.6$\tablefootmark{a}  \cr
NGC\,4579    & $ <1.6$      & $13.1 \pm 0.5$ & $5.0  \pm 0.2 $ & $2.3 \pm 0.3$ & $2    \pm 1$\tablefootmark{a}    \cr
NGC\,3169    & $ <0.24$     & $7.5  \pm 0.1$ & $2.8  \pm 0.2 $ & $1.2 \pm 0.1$ & $2.0  \pm 0.8$\tablefootmark{a}  \cr
NGC\,4261    & $ <0.27$     & $3.0  \pm 0.2$ & $2.0  \pm 0.2 $ & $0.7 \pm 0.1$ & $0.7  \pm 0.3$\tablefootmark{a}  \cr
M87          & $0.7 \pm 0.3$ & $4.9  \pm 0.3$ & $4.3  \pm 0.3 $ & $2.0 \pm 0.2$ & $0.7  \pm 0.3$\tablefootmark{a}  \cr
NGC\,4594    & $1.9 \pm 0.5$ & $8.5  \pm 0.2$ & $7.1  \pm 0.2 $ & $1.3 \pm 0.3$ & $2.4  \pm 0.6$\tablefootmark{a}  \cr
NGC\,404     & $ <0.21 $     & $2.5  \pm 0.1$ & $0.50 \pm 0.05$ & $0.6 \pm 0.1$ & $<1.0 $        \cr
\hline\\[-0.3cm]
3C\,273      & $3.6 \pm 0.3$ & $1.1  \pm 0.2$ & $4.2  \pm 0.3 $ & $2.6 \pm 0.5$ & $8.5  \pm 0.6$ \cr
\end{tabular}
\tablefoot{
\tablefoottext{a}{Line flux measured using a Gaussian function, to avoid possible contamination from the adjacent [\ion{Fe}{ii}]$_{26}$ line.}
}
\end{table*}

\subsection{Sub-arcsecond resolution imaging}\label{imaging}
The basis of this work is a high-angular resolution dataset that covers multiple spectral windows --\,radio, IR, optical/UV and X-rays\,-- for the inner few kiloparsecs of a sample of 9 nearby LLAGN. In particular, the use of AO in the near-IR with VLT/NaCo \citep{lenzen2003,rousset2003} and PISCES \citep{mccarthy2001} at the Large Binocular Telescope\footnote{The LBT is an international collaboration among institutions in the United States, Italy and Germany. LBT Corporation partners are: The University of Arizona on behalf of the Arizona university system; Istituto Nazionale di Astrofisica, Italy; LBT Beteiligungsgesellschaft, Germany, representing the Max-Planck Society, the Astrophysical Institute Potsdam, and Heidelberg University; The Ohio State University, and The Research Corporation, on behalf of The University of Notre Dame, University of Minnesota and University of Virginia.} (LBT) for the case of NGC\,404, and diffraction-limited imaging in the mid-IR with VLT/VISIR \citep{lagage2004}, allowed us to fill the gap in the $1$--$20\, \rm{\micron}$ region with genuine nuclear measurements, minimising the contamination from the underlying galaxy. For the optical/UV range, we used diffraction-limited observations from the \textit{Hubble Space Telescope} (\textit{HST}) scientific archive. Nuclear fluxes were extracted from the near-IR and optical images using aperture photometry of the unresolved component at the centre ($\lesssim 0\farcs1$), subtracting the local background around ($0\farcs2$--$0\farcs3$) in order to remove the starlight contribution from the host galaxy. The mid-IR photometry, mainly taken from \citet{reunanen2010} and \citet{asmus2014}, was measured in these works by fitting a point-spread function (PSF) and/or a Gaussian distribution to the nuclear unresolved component ($\lesssim 0\farcs4$). Measurements in the radio and X-ray ranges have been collected after an extensive and careful search in the literature. These correspond mainly to sub-arcsecond resolution observations ($< 1''$) acquired with the Very Large Array (VLA) and Very Long Baseline Interferometry (VLBI) in radio (1--10\,mas), and \textit{Chandra} ($\sim$\,2$''$), \textit{XMM}-Newton ($\sim$\,7$''$) and \textit{Integral} ($\sim$\,12$'$) at X-rays. Despite of the lower spatial resolution at X-rays, fluxes at high energies are generally dominated by the AGN due to the negligible contribution from stellar sources in this range. Nevertheless, at very low luminosities (L$_{\rm X} \lesssim 10^{40}$) the non-AGN contribution could still be an issue. In our sample, the faintest galaxies in X-rays are NGC\,1097, M87, Sombrero, and NGC\,404 (see Table\,\ref{tab_sample}). In the case of NGC\,1097, the superior spatial resolution of \textit{Chandra} is able to separate the active nucleus from the circumnuclear star-forming ring, thus avoiding the contamination from the latter \citep{nemmen2006,mezcua2015}. Early-type galaxies such as M87 and NGC\,4594 are less likely affected by contamination from star formation activity \citep{pellegrini2003}. Finally, X-ray variability has been detected in M87 and NGC\,404 \citep{harris2009,binder2011}, which is indicative of AGN activity. Therefore, the X-ray measurements compiled for this work can be consistently compared with sub-arcsecond measurements at other wavelengths. The characteristics of this dataset allow us to build, for each LLAGN in the sample, a complete continuum spectra of the same physical region, very well sampled over a wide range in wavelength (see Fig.\,\ref{fig_seds}). Additionally, a low-spatial resolution continuum based on apertures larger than few arcsec was also built using the NED database\footnote{\url{http://ned.ipac.caltech.edu}} and measurements published in the literature. The low-spatial resolution data are complementary to the sub-arcsecond continuum spectra, as they permit to identify those spectral ranges in which the emission is potentially dominated by the host galaxy, contaminating the nuclear spectrum. The complete multi-wavelength catalogue with both the wide aperture and the sub-arcsec nuclear fluxes for the 9 LLAGN in the sample are included as part of the online materials associated with this publication.

\subsection{Mid-IR spectroscopy}\label{irspec}
\begin{figure*}
\centering
\includegraphics[width = 0.45\textwidth]{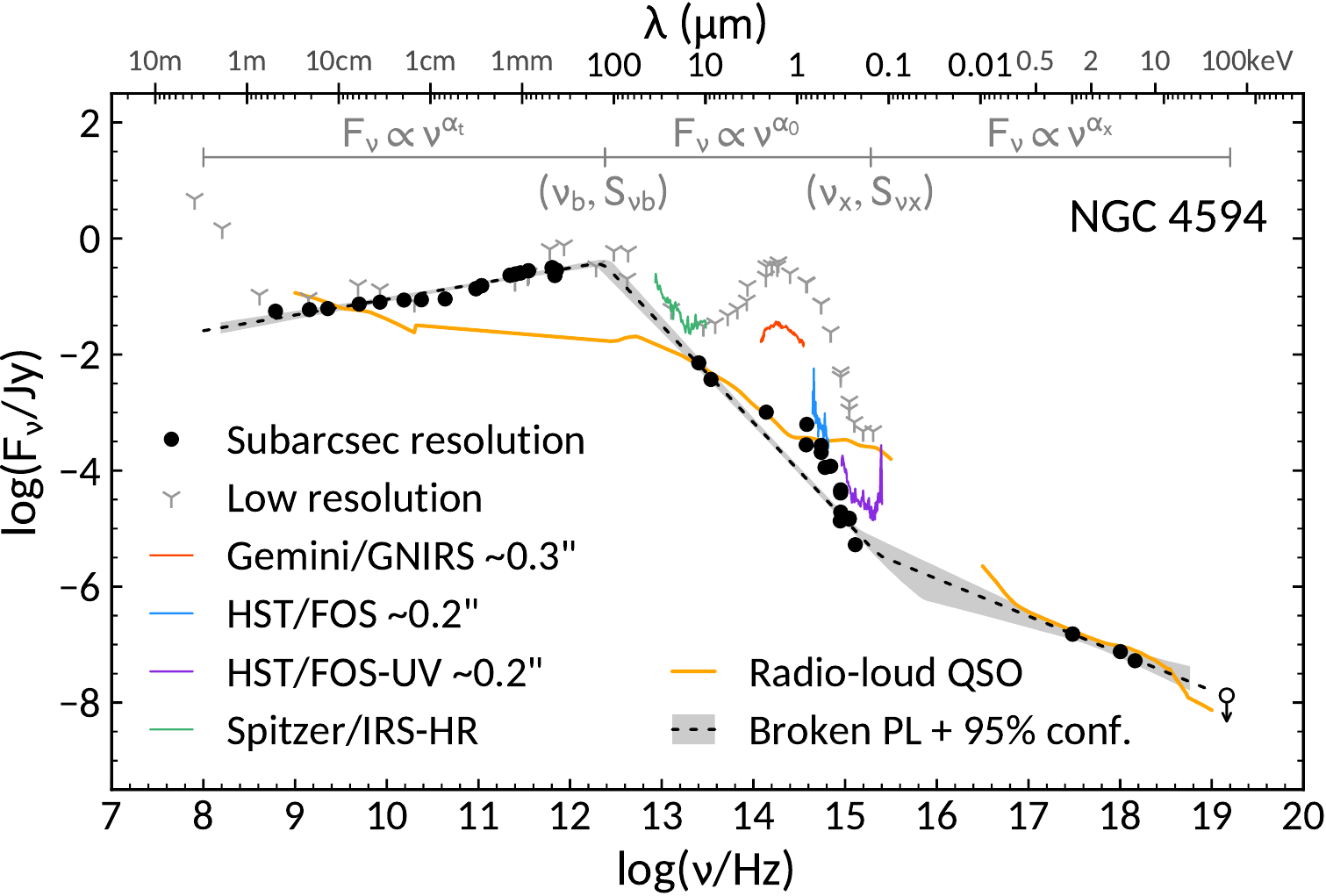}~
\includegraphics[width = 0.45\textwidth]{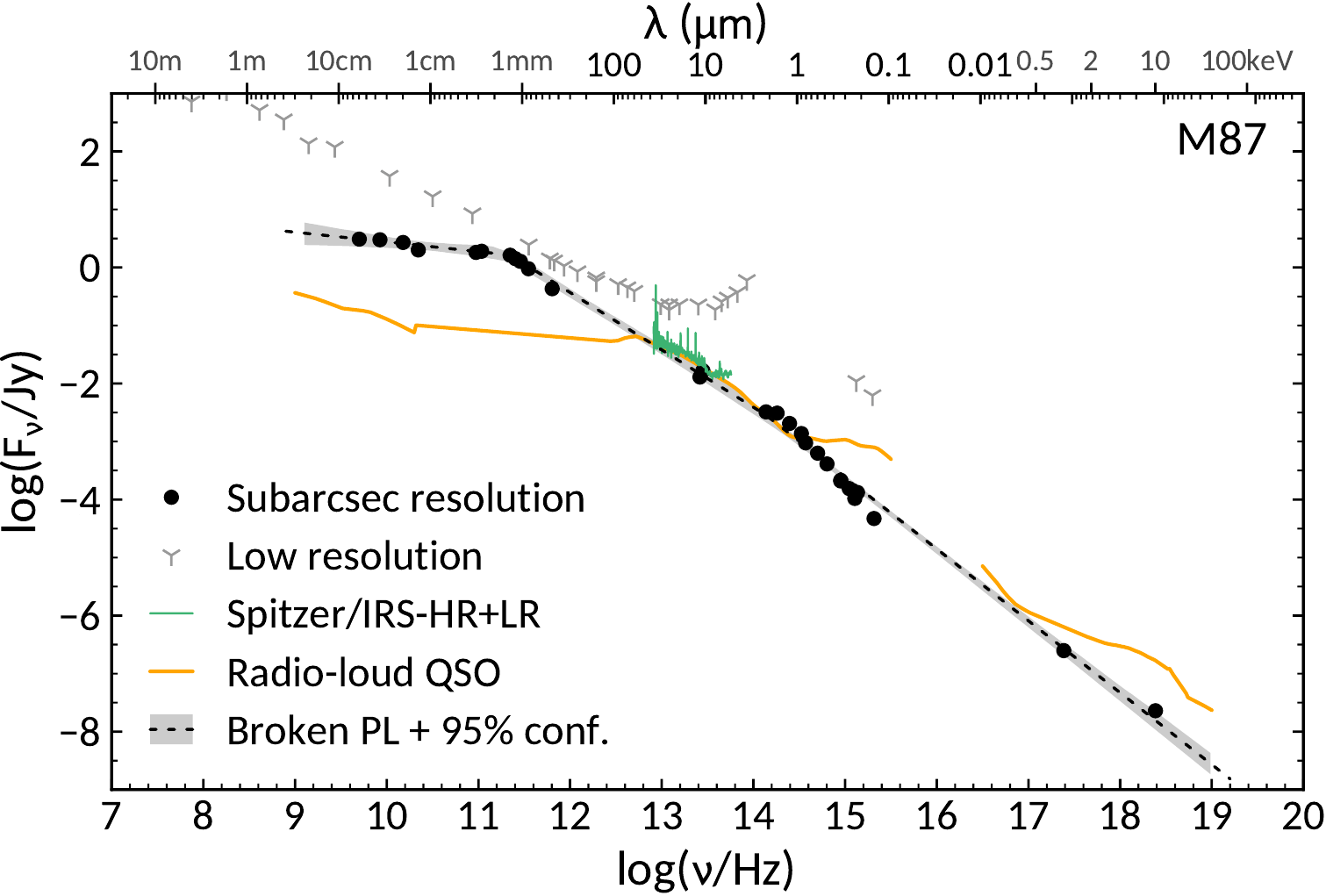}\\
\includegraphics[width = 0.45\textwidth]{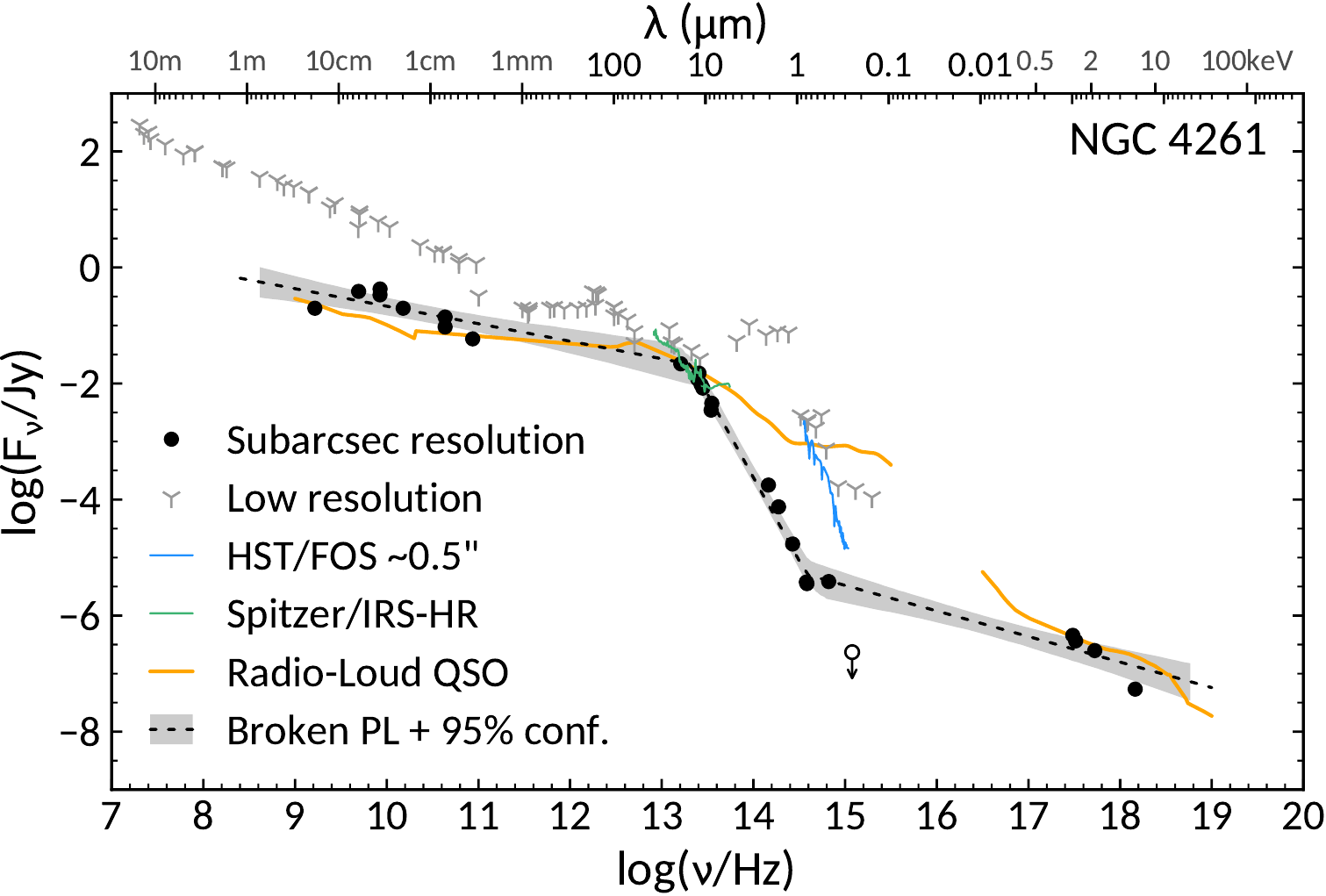}~
\includegraphics[width = 0.45\textwidth]{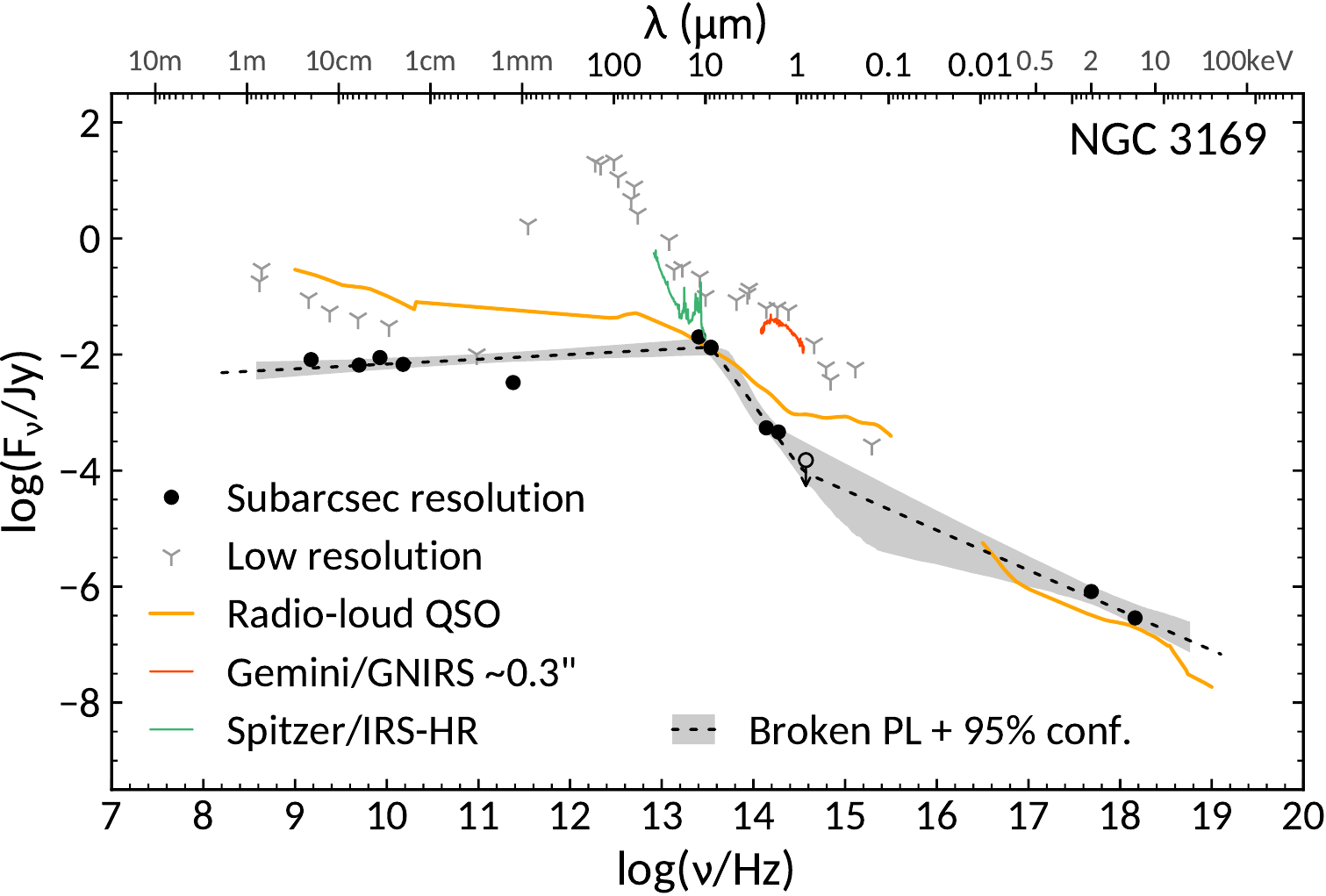}\\
\includegraphics[width = 0.45\textwidth]{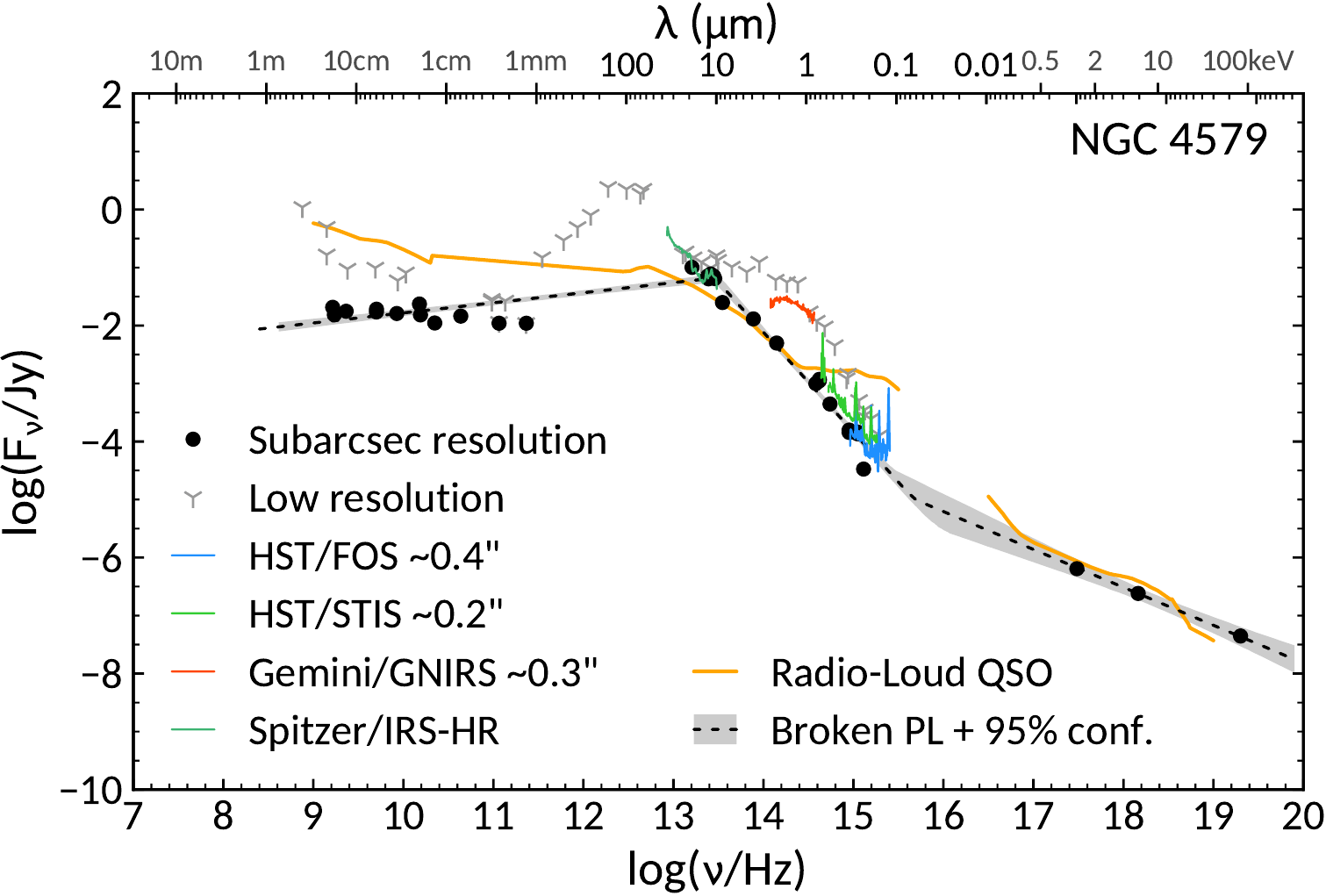}~
\includegraphics[width = 0.45\textwidth]{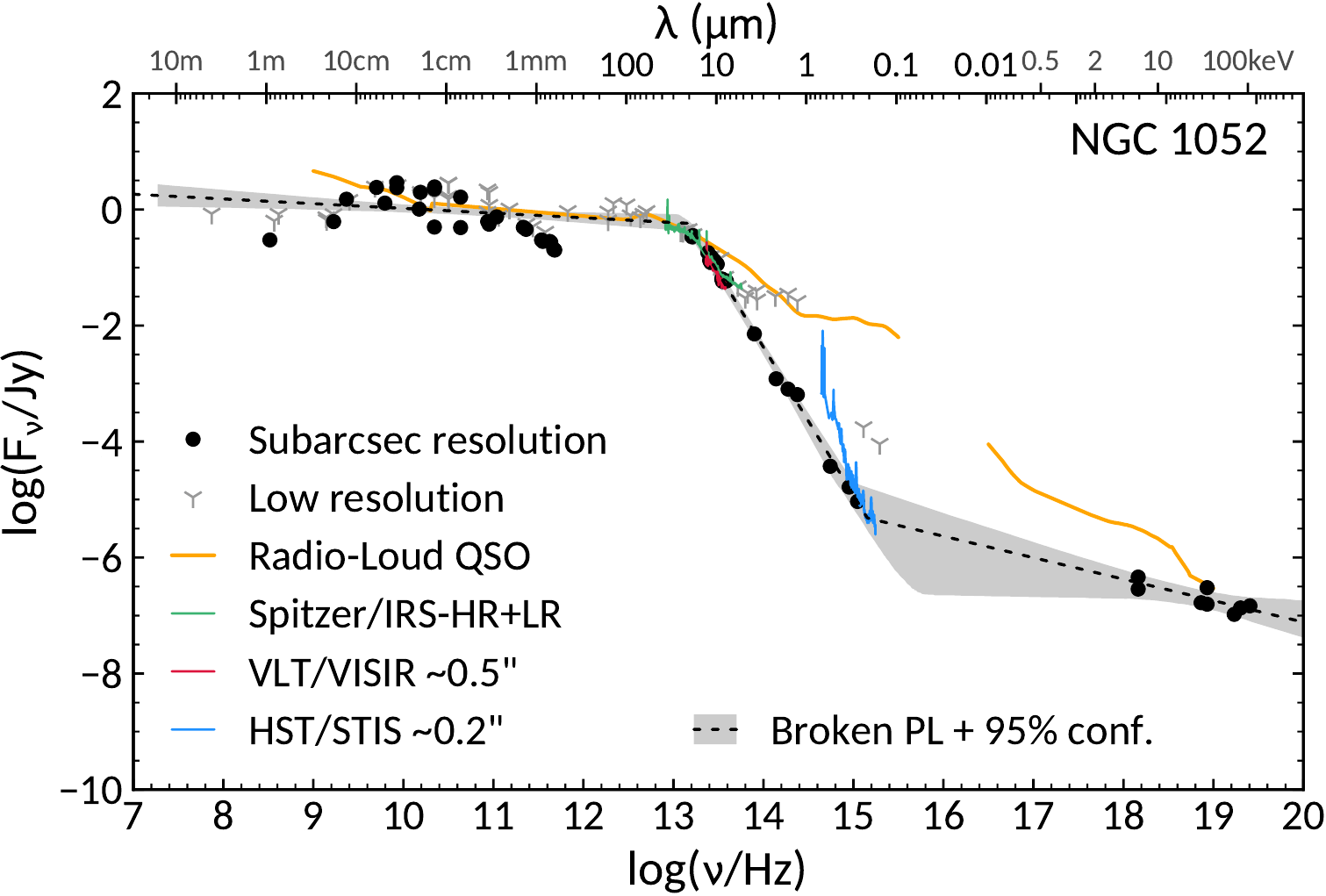}\\
\includegraphics[width = 0.45\textwidth]{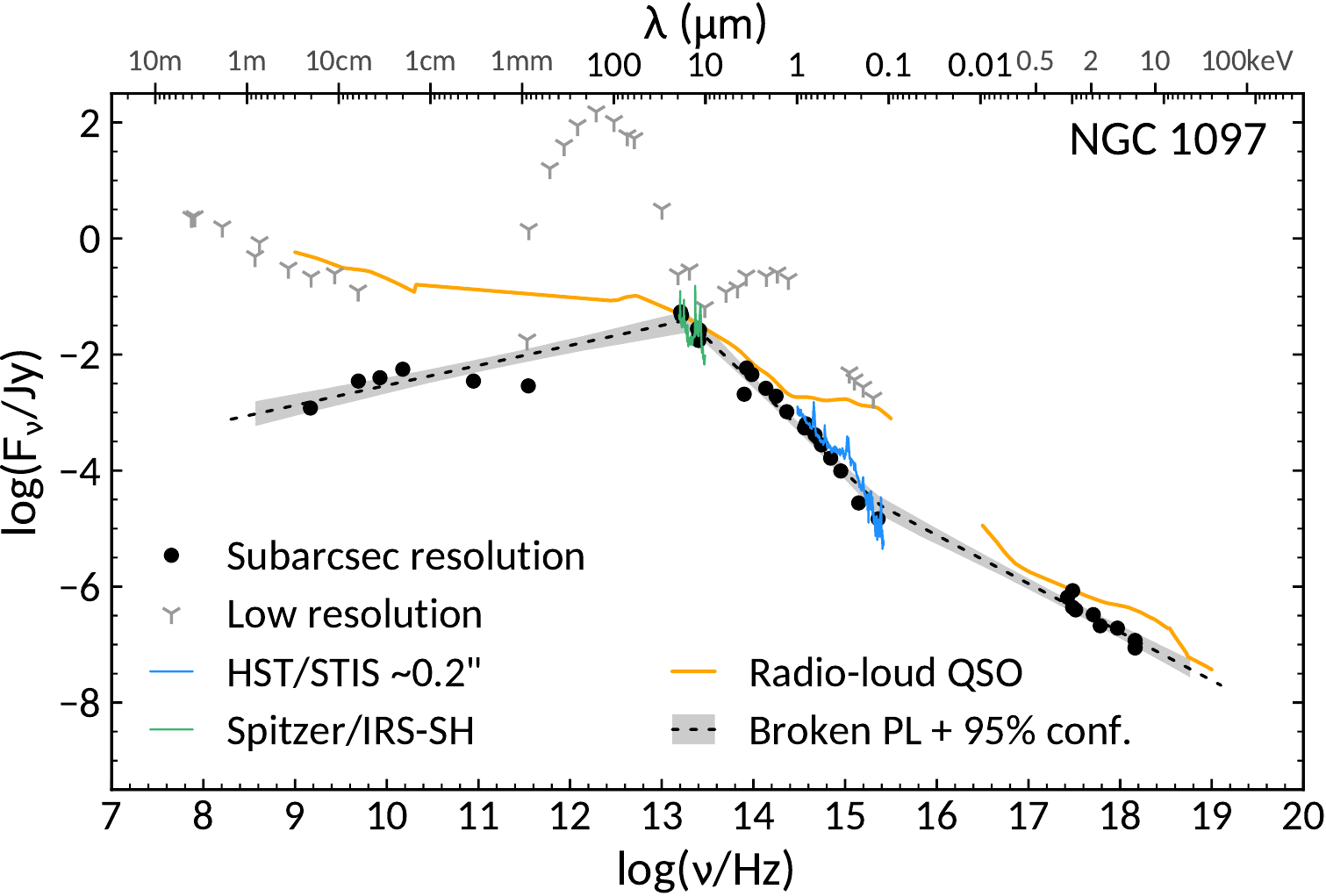}~
\includegraphics[width = 0.45\textwidth]{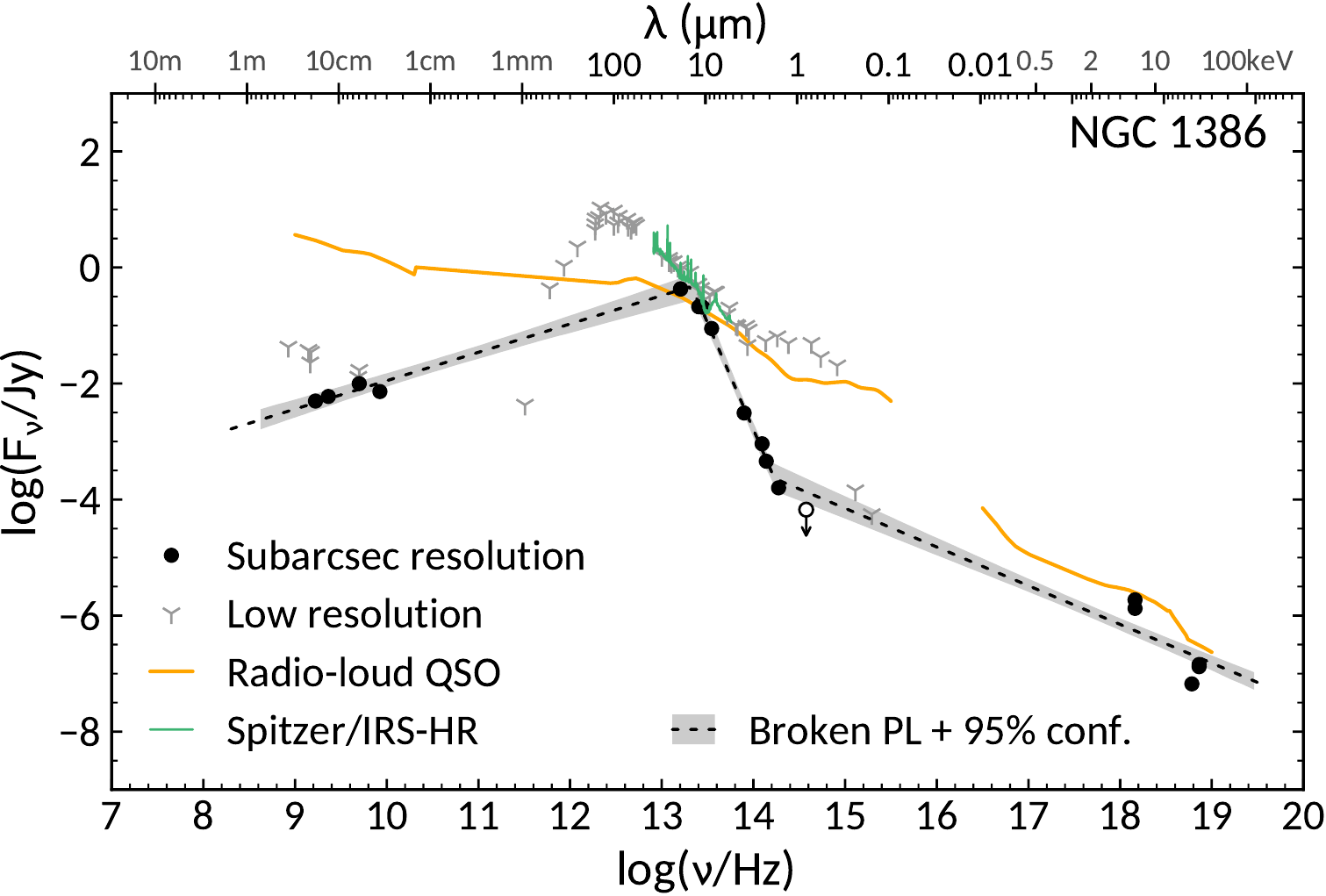}
\vspace{-0.1cm}
\caption{Continuum spectra based on sub-arcsecond (black dots) and large aperture data (grey spikes) for the sample of LLAGN, sorted by increasing $\lambda_{\rm Edd}$ (Table\,\ref{tab_sample}). A three piece broken power law has been fitted to the sub-arcsecond flux density distribution (dashed-line), intended as a parametrisation of the compact jet emission. We added archival spectra from \textit{Spitzer}/IRS in the mid-IR (in green) and \textit{HST}/FOS and STIS in the optical/UV (in blue and purple). Gemini/GNIRS near-IR spectra (in orange) from \citet{mason2015} and VLT/VISIR mid-IR spectra for NGC\,1052 (in red; \citealt{jafo2019}) are also shown. The radio-loud quasar template from \citet{1994ApJS...95....1E} is also shown for compasigon (orange line).}\label{fig_seds}
\end{figure*}

\begin{figure*}[hbtp!!!]
\centering
\includegraphics[width = 0.5\textwidth]{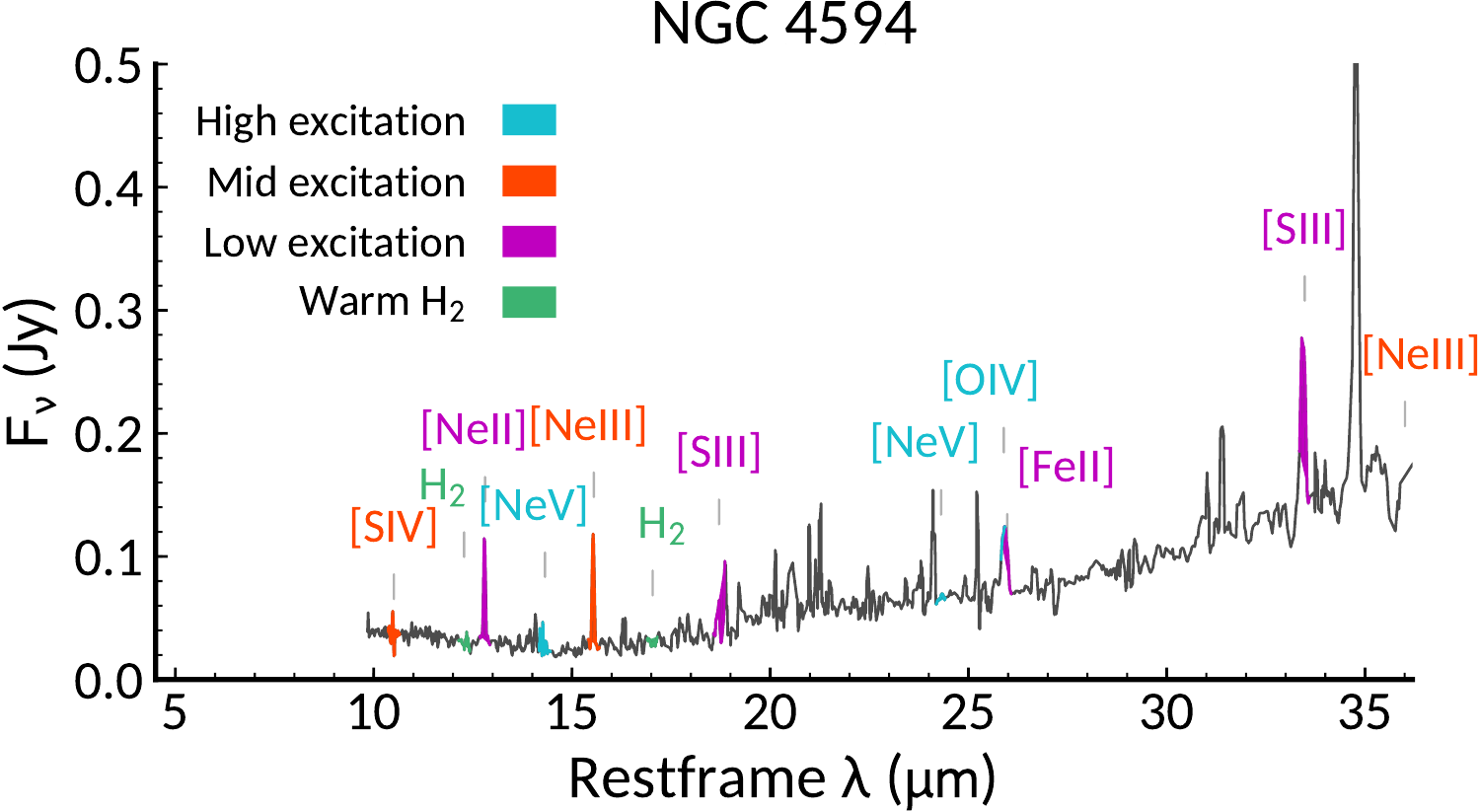}~
\includegraphics[width = 0.5\textwidth]{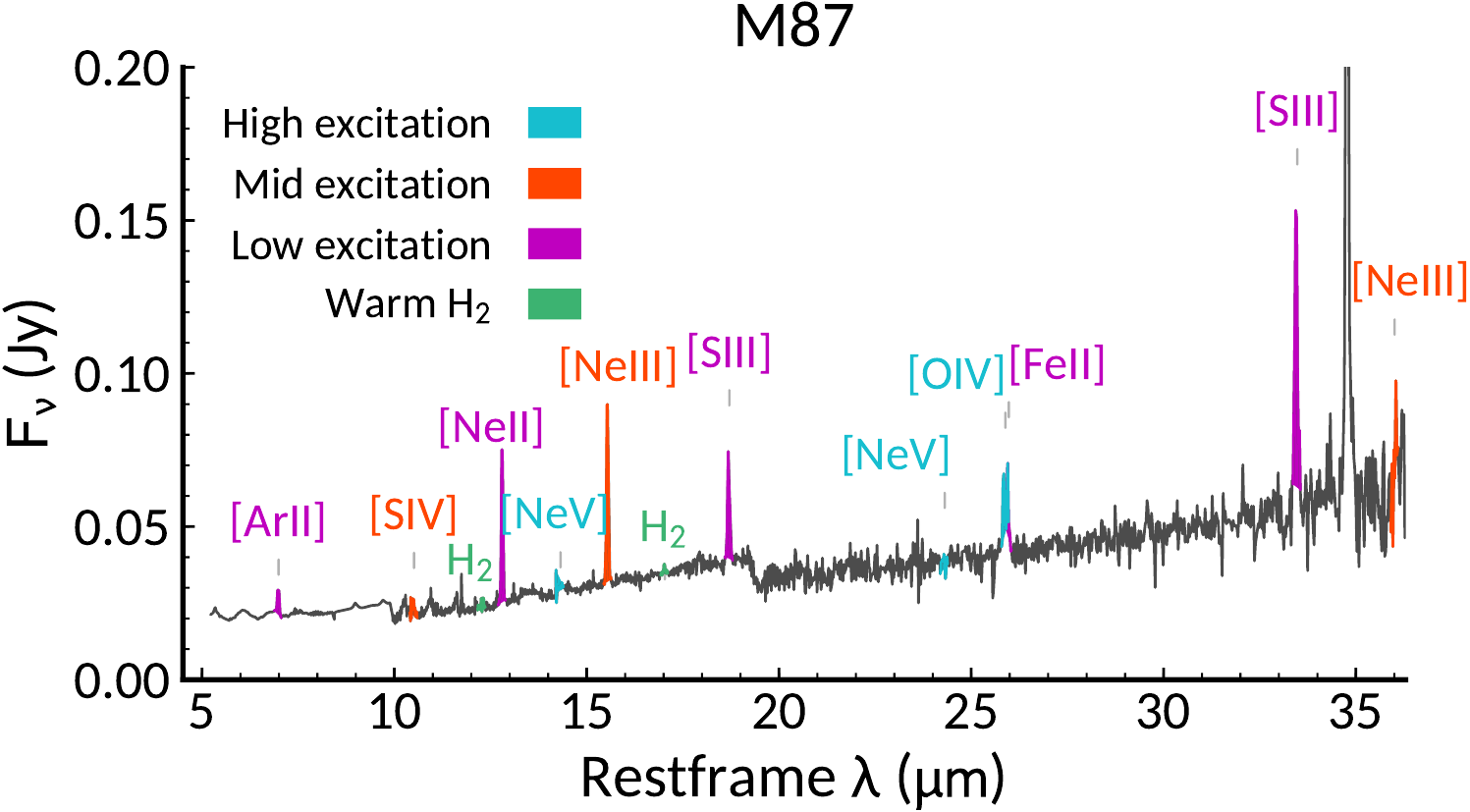}\\[0.27cm]
\includegraphics[width = 0.5\textwidth]{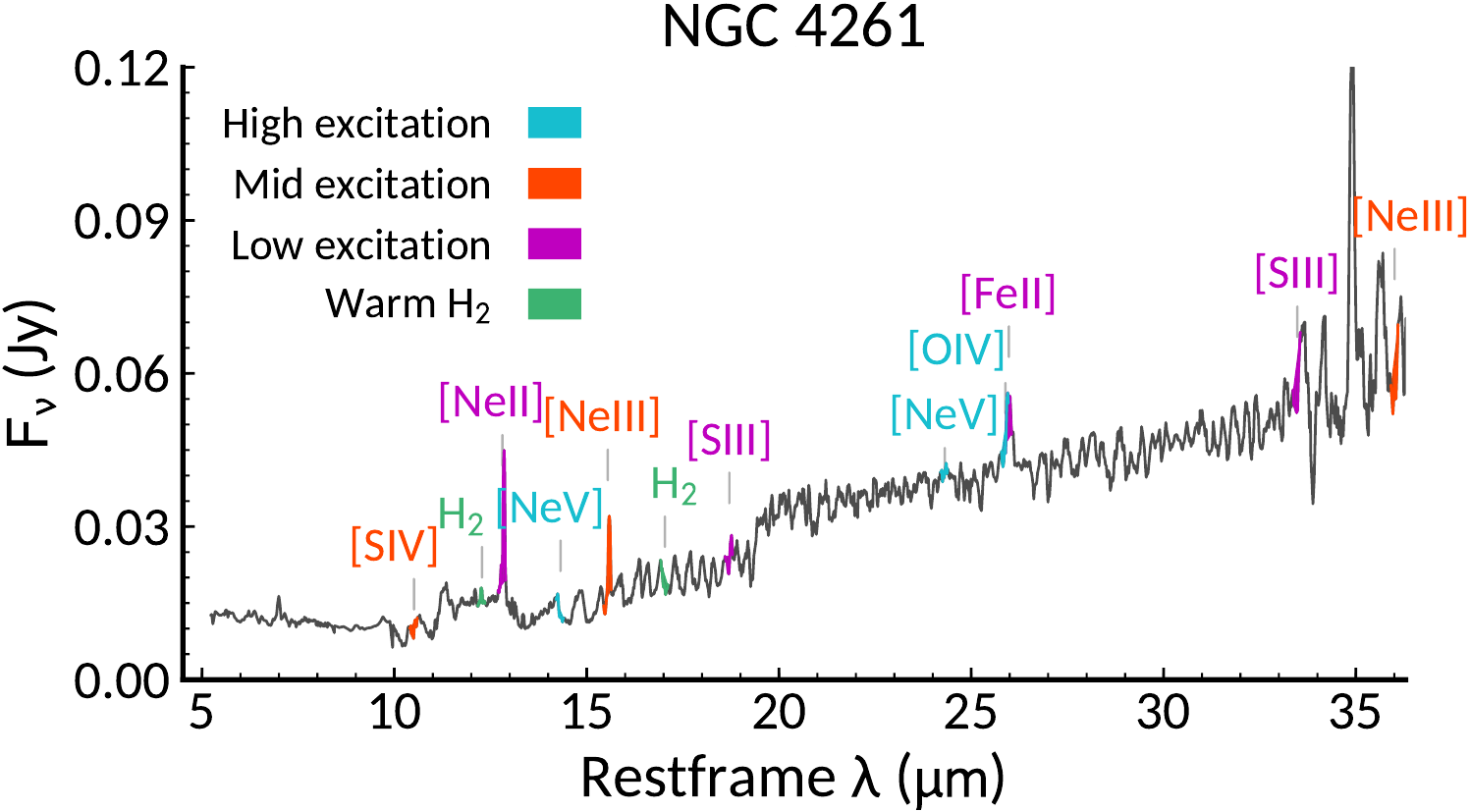}~
\includegraphics[width = 0.5\textwidth]{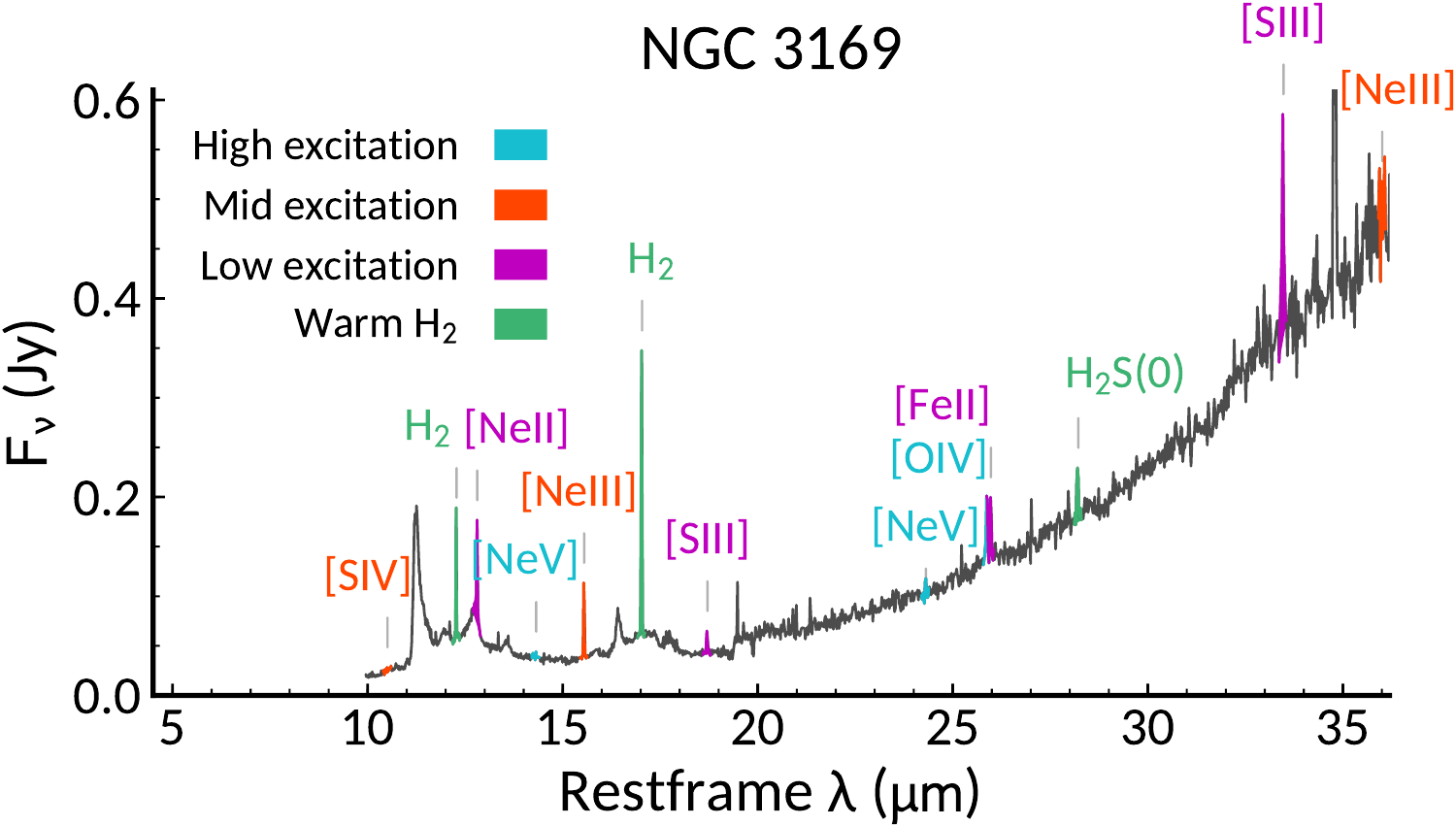}\\[0.27cm]
\includegraphics[width = 0.5\textwidth]{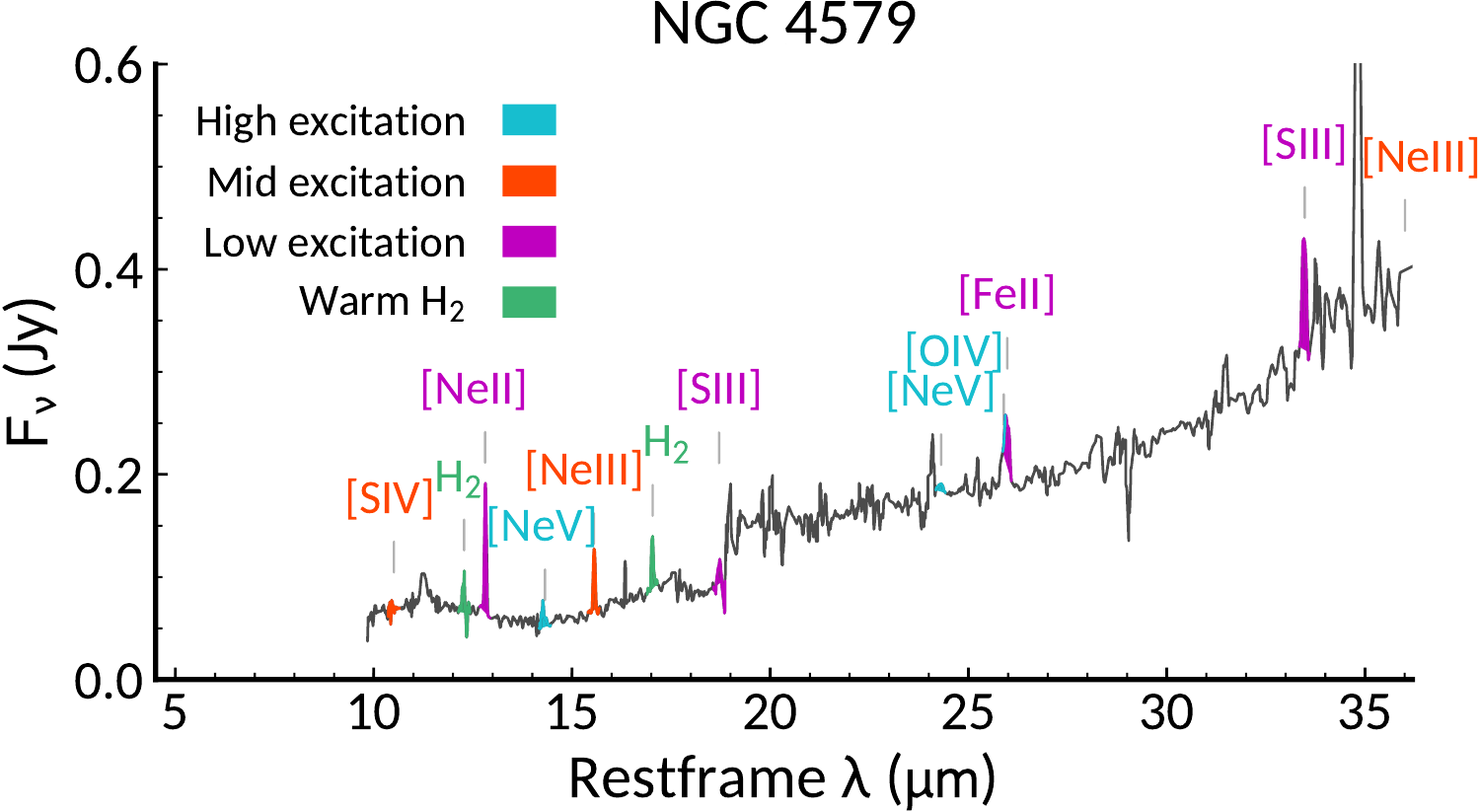}~
\includegraphics[width = 0.5\textwidth]{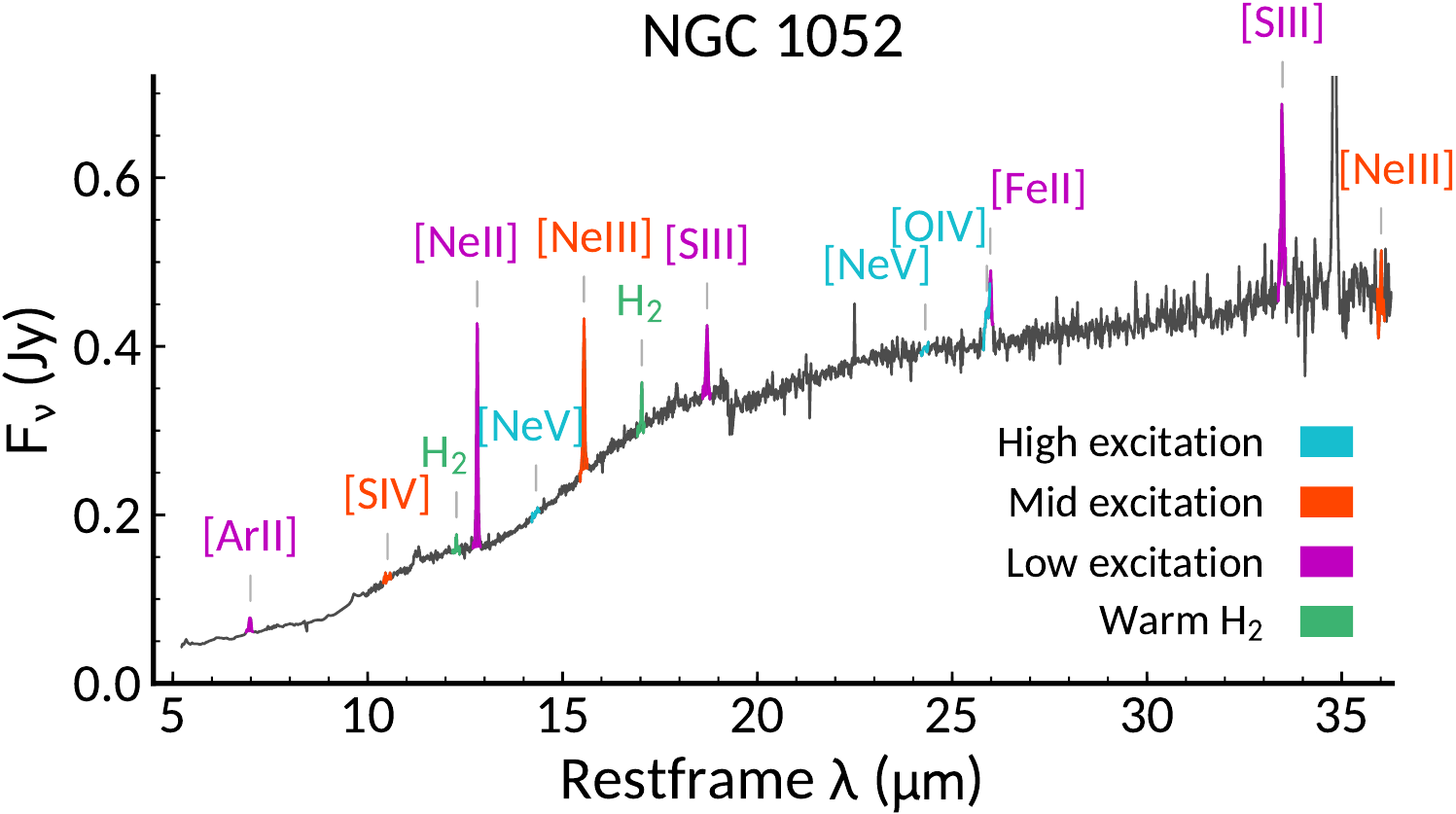}\\[0.27cm]
\includegraphics[width = 0.5\textwidth]{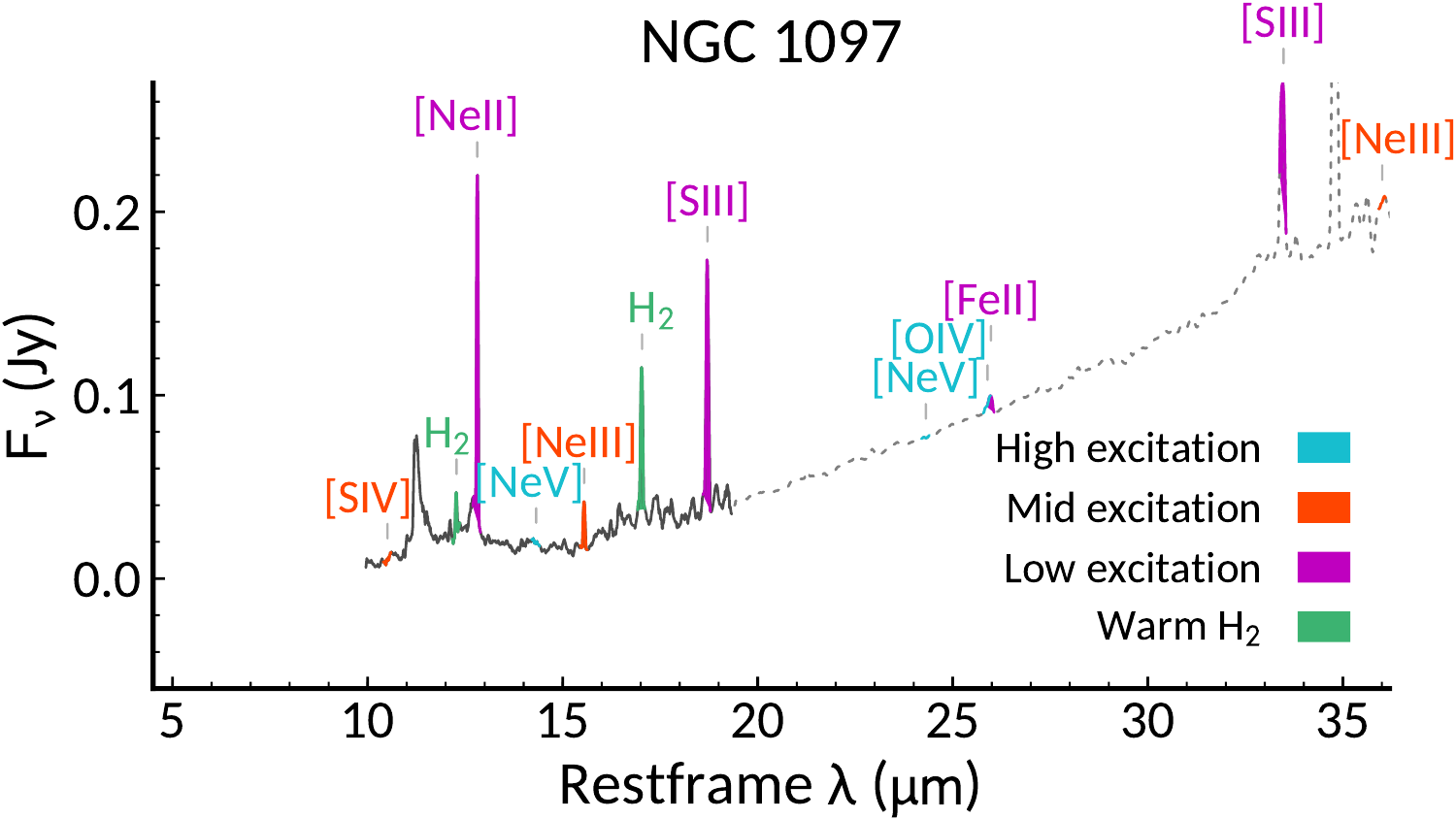}~
\includegraphics[width = 0.5\textwidth]{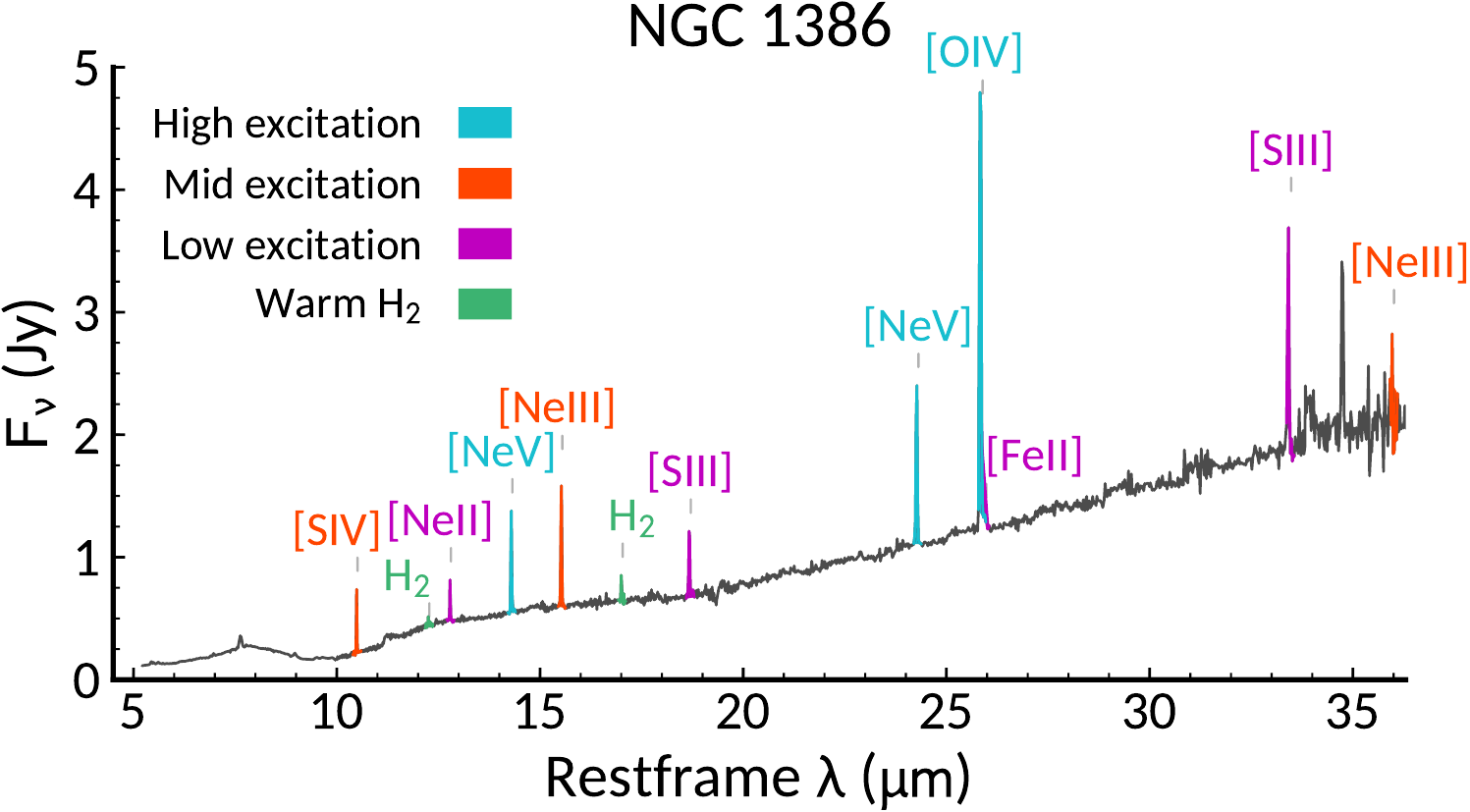}\\[0.1cm]
\caption{\textit{Spitzer}/IRS mid-IR spectra for the sample of LLAGN obtained from the CASSIS database \citep{lebouteiller2011,lebouteiller2015}, including both the low- and the high-resolution modules. The main nebular and molecular gas lines are indicated, except for the [\ion{Si}{ii}]$_{34.8}$ line which typically exceeds the flux range shown in the figure. The objects are sorted by increasing Eddington rate ($\lambda_{\rm Edd}$; see Table\,\ref{tab_sample}). In the case of NGC\,1097, the continuum in the long-high module (dotted line) has been scaled to match the short-high module at $19\, \rm{\micron}$, due to contamination from the star forming ring in this galaxy.}\label{fig_mirspec}
\end{figure*}

\begin{figure*}
 \centering
  \includegraphics[width=0.5\textwidth]{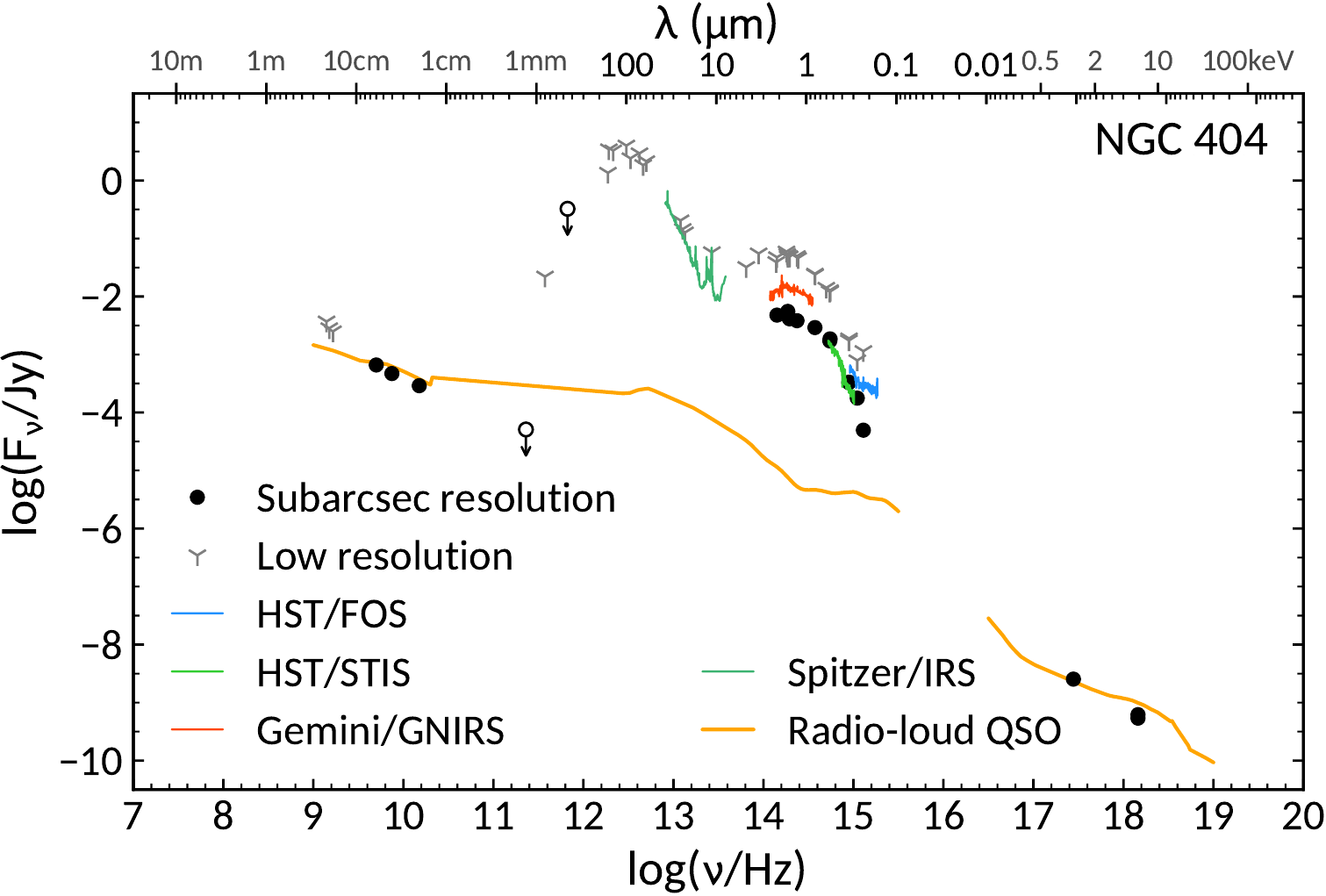}~
  \includegraphics[width=0.5\textwidth]{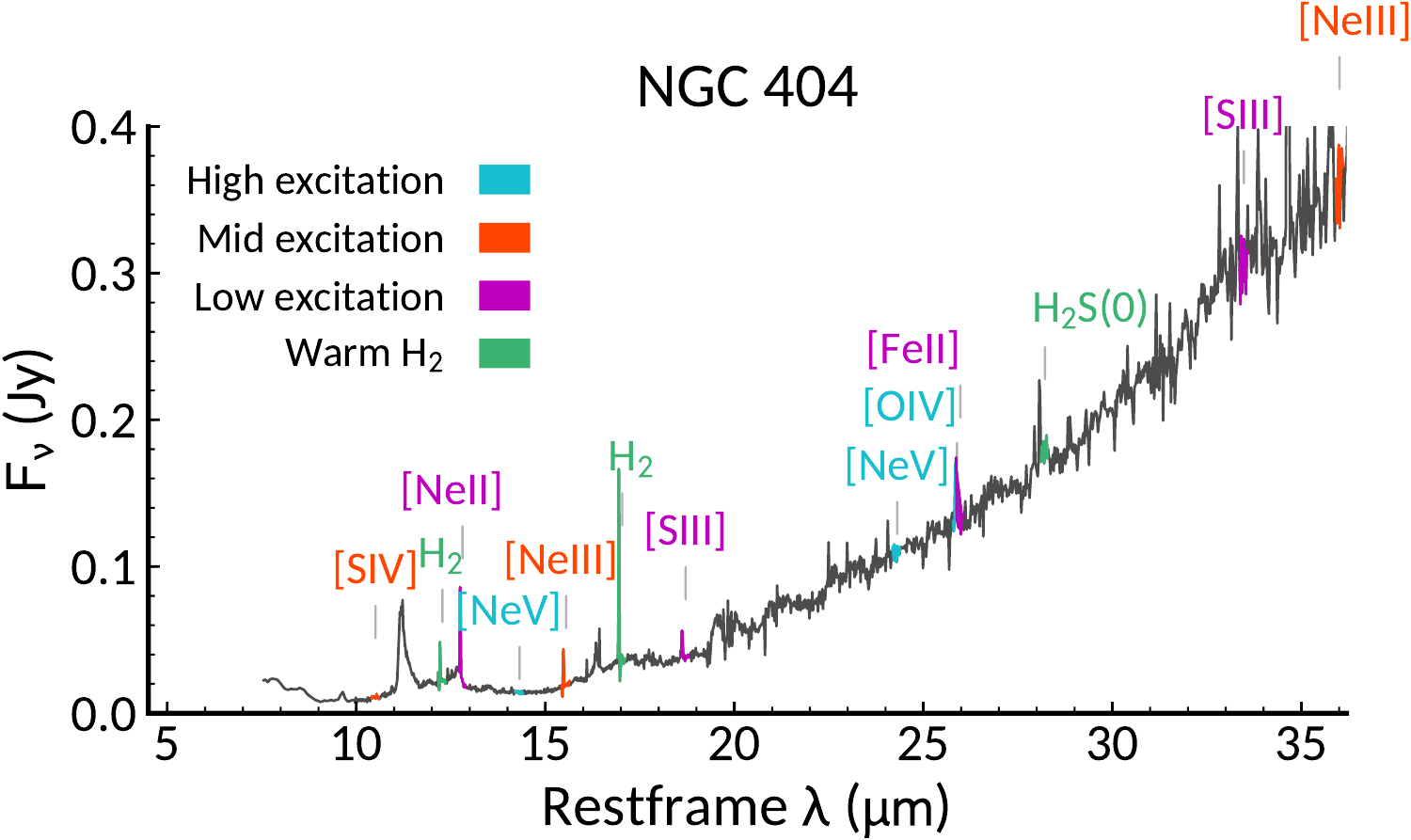}
  \caption{Continuum flux spectral density distribution for the nucleus of NGC\,404 (\textit{left}), and mid-IR spectrum from \textit{Spitzer}/IRS (\textit{right}). The latter combines the low- ($\lesssim 9.9\, \rm{\micron}$) and high-spectral resolution datasets. Symbols as per Figs.\,\ref{fig_seds} and \ref{fig_mirspec}.}\label{fig_n404}
\end{figure*}

In order to extend our knowledge on the shape of the continuum emission in LLAGN, we collected \textit{Spitzer}/IRS spectra for our sample of galaxies acquired with both the low- ($R \sim 60$--$120$, $\Delta v \sim 2500$--$5000\, \rm{km\,s^{-1}}$) and the high-spectral resolution modes ($R = 600$, $\Delta v = 500\, \rm{km\,s^{-1}}$) from the CASSIS database (Combined Atlas of Sources with \textit{Spitzer} \textsc{Irs} Spectra; \citealt{lebouteiller2011,lebouteiller2015}). These mid-IR spectra have been reduced, calibrated, and extracted using a dedicated pipeline to perform an optimal PSF extraction for point-like sources, removing a significant fraction of the extended emission typically included within the relatively wide IRS aperture for the short-high module ($4\farcs7 \times 11\farcs3$ in the $9.9$--$19.5\, \rm{\micron}$ range), and the long-high module ($11\farcs1 \times 22\farcs3$ in the $18.8$--$37.1\, \rm{\micron}$ range). This results in a smoother and a more consistent continuum level at $\sim 19\, \rm{\micron}$ and a significantly reduced contribution from PAH features and low-ionisation nebular lines, that might originate in star forming regions in the innermost few kpc of these galaxies \citep[e.g. fig.\,2 in][]{jafo2018}. NGC\,4579 and NGC\,4594 were not included in the CASSIS database, the calibrated mid-IR spectra in these cases were gathered from the \textit{Spitzer} Heritage Archive\footnote{\url{https://sha.ipac.caltech.edu/applications/Spitzer/SHA/}}, using the full slit aperture to derive the integrated spectra. The mid-IR spectra for all the LLAGN in the sample are compared with the photometric measurements in Figs.\,\ref{fig_seds} and \ref{fig_n404} (left panel), while the detailed spectra in the $5$--$35\, \rm{\micron}$ range is shown in Figs.\,\ref{fig_mirspec} and \ref{fig_n404} (right panel). A Wiener filter has been applied to the mid-IR spectra of M87, NGC\,4261, NGC\,4579, and NGC\,4594 in order to increase the signal-to-noise for the faintest sources. A filter width of $\sim 500\, \rm{km\,s^{-1}}$ was used to smooth flux values below $3 \times$\,\textsc{rms}, while the emission lines remained unaffected. Line fluxes were measured using our own \textit{Python}\footnote{\url{https://www.python.org}} routines by integrating the line profile and subtracting the continuum level, which was determined from a 1D polynomial fit to the flux density distribution of the spectral elements located at both sides of the line. For the case of the [\ion{O}{iv}]$_{25.9}$ line, a Gaussian function was used in most cases to measure the line flux in order to avoid possible contamination from the adjacent [\ion{Fe}{ii}]$_{26}$ line (Table\,\ref{table_IR_lines}). The measured fluxes for the main nebular lines in the mid-IR range are shown in Table\,\ref{table_IR_lines}. Previous measurements of these lines published in the literature were based on the full integration of the IRS slits, and therefore exposed to heavier contamination of the low-ionisation lines by star formation activity, especially [\ion{Ne}{ii}]$_{12.8}$ and [\ion{S}{iii}]$_{18.7}$. The new values provided here are based on the optimally extracted spectra and thus are less affected by this problem.

However, the IRS long-wavelength spectrum in NGC\,1097 is still affected by strong contamination from extended emission (see Fig.\,\ref{fig_mirspec}). This is caused by the contribution of the $1\, \rm{kpc}$ diameter star formation ring in this galaxy, which outshines the active nucleus in the mid-IR and it fills partially the long-wavelength slit, and possibly also the short-wavelength module (fig.\,1 in \citealt{bernard-salas2009}, \citealt{prieto2019}). This problem will be further discussed in Section\,\ref{hidden}.

%\subsection{Variability}\label{varia}

%\textcolor{Turquoise}{Variability in NGC\,404? \citep{2017ApJ...836..237N}, NGC\,4594 (optical/UV), NGC\,1097? (optical, \textit{L'}-band, and at $1\, \rm{mm}$), M87, NGC\,1052 (mm bursts).}

\section{Results}\label{results}

The first step in our analysis is to build the high-angular resolution spectrum for the nine LLAGN in the sample, obtained from the sub-arcsecond resolution images and the radio and X-ray measurements collected from the literature (Section\,\ref{shape}). Because the ionising continuum in the EUV range cannot be directly observed, we follow an indirect approach to infer the EUV continuum shape by comparing the observed ratios of the mid-IR nebular lines with predictions from photo-ionisation simulations (Section\,\ref{cloudy}).

\subsection{The continuum emission at high-angular resolution}\label{shape}
The nuclear fluxes across the electromagnetic spectrum obtained from sub-arcsecond aperture measurements are shown in Figs.\,\ref{fig_seds} and \ref{fig_n404} for the 9 LLAGN in our sample (black dots). For comparison, we include the fluxes measured with large apertures --\,typically a few arcsecs to a few arcmins\,-- as grey spikes. These are significantly brighter than sub-arcsecond fluxes in those spectral ranges where the host galaxy light dominates the emission, usually in the optical due to the stellar component and in the mid- to far-IR due to dust heated by star formation activity. This is particularly evident in the case of NGC\,1097, where large aperture fluxes are dominated by the starburst ring from radio to optical/UV wavelengths. The nucleus, fainter by more than an order of magnitude, is revealed only by using high-angular resolution data or at high energies ($\gtrsim 0.5\, \rm{keV}$). On the other hand, in the case of NGC\,1052, the comparison between high- and low-angular resolution fluxes reveals a steep nuclear continuum that completely overcomes the host galaxy emission longwards of $\gtrsim 20\, \rm{\micron}$.

The lack of a big blue bump in LLAGN is evident when these are compared to the average radio-loud quasar from \citet{1994ApJS...95....1E}, shown as an orange-solid line scaled to the $\sim 10\, \rm{\micron}$ sub-arcsecond flux in Figs.\,\ref{fig_seds} and \ref{fig_n404}. In contrast, the main characteristic that stands out in the sub-arcsecond flux density distributions of LLAGN is a very steep continuum decrease from the IR to the UV range, sharply defining a power-law relation over $\sim 2$ orders of magnitude in wavelength ($-3.7 \lesssim \alpha_0 \lesssim -1.3$; $S_\nu \propto \nu^{\alpha_0}$). This behaviour, already noticed by \citet{ho1999} for a sample of LINERs based on optical \textit{HST} imaging, extends over the whole near- to mid-IR range when high-angular resolution data are considered. In some cases the dust obscuration blocks the nuclear optical/UV light, thus the optically thin continuum can only be traced in the mid- to near-IR range (e.g. NGC\,1386, NGC\,3169). Only the Sombrero galaxy (NGC\,4594), which has the lowest luminosity in Eddington units (Table\,\ref{tab_sample}), shows a thermal-like bump peaking at $\sim 1\, \rm{\micron}$ (upper-left panel in Fig.\,\ref{fig_seds}). Although this wavelength is close to the maximum contribution from the underlying stellar population (grey spikes), the nucleus of Sombrero presents a mild variability in the \textit{HST} data, that is its flux in the \textit{V} and \textit{I} bands increased by a factor $\sim 1.5$--$2$ in the 1994--1996 period (see flux tables included in the online materials associated with this publication, in agreement with the variability seen in the UV range \citep{maoz2005}. As it is shown in this work, this bump is is well fit by a cold and/or truncated accretion disc ($\lesssim 3000\, \rm{K}$), still an underlying power-law continuum is required to explain the rise of the flux in the $10$--$100\, \rm{\micron}$ range and possibly also in the far-UV above $\gtrsim 10^{15}\, \rm{Hz}$ (dashed-black line in the upper-left panel of Fig.\,\ref{fig_seds}).

The case of NGC\,404 is different from the rest of LLAGN in our sample, since the nuclear continuum does not follow a power-law behaviour in the IR-to-optical/UV. Its sub-arcsecond flux density distribution is rather flat longwards of $\sim 0.8\, \rm{\micron}$ with a steep decrease at bluer wavelengths, a similar shape to that observed for the stellar component in the large aperture measurements (left panel in Fig.\,\ref{fig_n404}). This is the closest galaxy in the sample hosting also the lowest mass BH ($1.5 \times 10^5\, \rm{M_\odot}$; \citealt{nguyen2017}). Its active nucleus have been revealed at X-ray and radio wavelengths \citep{binder2011,nyland2012}, although the relative radio-to-optical and X-ray-to-optical intensities appear to be significantly lower when compared to other nuclei in the sample, suggesting that the nuclear emission in NGC\,404 may still be contaminated by the starlight even in sub-arcsecond resolution data. Nevertheless, the nucleus shows feeble variability in the optical/UV, changing its flux by $\sim 20$\,\% over a 15 year period \citep{nguyen2017}. Hereafter we mainly focus on the eight LLAGN where the nuclear emission has been distinctively isolated from the host galaxy, while NGC\,404 will require further observations at higher angular resolution in order to separate the nuclear component.
\begin{table*}
%\footnotesize
\centering
\setlength{\tabcolsep}{3.pt}
\caption{Broken power-law fits to the continuum spectra shown in Fig.\,\ref{fig_seds}. The columns correspond to the galaxy name, the spectral indices of the radio ($\alpha_{\rm t}$), IR-to-optical/UV ($\alpha_0$) and X-ray components ($\alpha_{\rm x}$), the turnover frequencies ($\nu_{\rm b}$ and $\nu_{\rm x}$) and their associated fluxes ($S_{\rm \nu b}$ and $S_{\rm \nu x}$). The different spectral components of the broken power-law distribution used to fit the continuum spectra are indicated for the case of the Sombrero galaxy in the upper left panel of Fig.\, \ref{fig_seds}.}\label{tab_fit}
\begin{tabular}{lccccccc}
\bf Name & \bf $\alpha_{\rm t}$ & \bf $\alpha_0$ & \bf $\alpha_{\rm x}$ & \bf $\nu_{\rm b}$ & \bf $\nu_{\rm x}$ & $S_{\rm \nu b}$ & $S_{\rm \nu x}$ \\
         &                &                &                &  [$10^{12}\, \rm{Hz}$] & [$10^{15}\, \rm{Hz}$] &   [mJy] & [$10^{-3}\,$mJy] \\
\hline\\[-0.3cm]
NGC\,4594 & $ 0.26 \pm 0.02$ & $ -1.68 \pm 0.01$& $-0.7 \pm 0.1$ & $2.4 \pm 0.3$  & $1.9  \pm 0.9$  & $380 \pm 30$  & $5 \pm 5$\\
M87       & $-0.17 \pm 0.06$ & $-0.97 \pm 0.08$ & $-1.2  \pm 0.2$  & $0.21\pm 0.03$ & $0.14 \pm 0.05$ & $1600 \pm 200$& $(3 \pm 1) \times 10^3$  \\
NGC\,4261 & $-0.31 \pm 0.05$ & $-2.86 \pm 0.06$ & $-0.44 \pm 0.05$ & $22 \pm 4$     & $0.384\pm 0.009$& $19  \pm 8$   & $5 \pm 2$ \\
NGC\,3169 & $0.08 \pm 0.02$  & $-2.12 \pm 0.02$ & $-0.73 \pm 0.08$ & $34.4\pm 0.4$  & $0.3  \pm 0.1$  & $14  \pm 2$   & $200 \pm 100$ \\
NGC\,4579 & $0.17 \pm 0.02$  & $-1.78 \pm 0.07$ & $-0.65 \pm 0.09$ & $31  \pm 7$    & $4    \pm 2$    & $64  \pm 6$   & $11  \pm 8$ \\
NGC\,1052 & $-0.09 \pm 0.03$ & $-2.65 \pm 0.08$ & $-0.44 \pm 0.06$ & $15  \pm 2$    & $0.9  \pm 0.3$  & $500 \pm 100$ & $11 \pm 7$ \\
NGC\,1097 & $0.34 \pm 0.04$  & $-1.59 \pm 0.04$ & $-0.80 \pm 0.04$ & $21  \pm 3$    & $2.3  \pm 0.1$\tablefootmark{a} & $41 \pm 8$ & $23 \pm 4$ \\
NGC\,1386 & $0.47 \pm 0.05$  & $-3.71 \pm 0.08$ & $-0.67 \pm 0.03$ & $22  \pm 2$    & $0.17 \pm 0.02$ & $400 \pm 100$ & $240 \pm 80$ \\
\hline\\[-0.3cm]
3C\,273 & $ -0.58 \pm 0.03$ & $-1.60 \pm 0.04$ & $-0.60 \pm 0.04$ & $(6.2 \pm 0.4) \times 10^3$ & $(1.6 \pm 0.2) \times 10^2$ & $2.9 \pm 0.1$ & $16.8 \pm 0.3$
\end{tabular}
\tablefoot{
\tablefoottext{a}{The error of $\nu_{\rm x}$ corresponds to the nearest filter bandwidth at this frequency, since the best-fit error was considerably smaller.}
}
\end{table*}

Low-luminosity AGN are known for being radio-loud sources \citep{terashima2003,nagar2005,ho2008}, and this is also the case for all the objects in our sample, showing a relatively bright and flat spectrum at radio wavelengths ($-0.3 \lesssim \alpha_{\rm t} \lesssim 0.5$). Moreover, all the sources other than NGC\,404 show a flat-to-inverted flux density distribution, that is a flat trend at radio wavelengths followed by the aforementioned power-law steep continuum in the IR-to-optical/UV range. This is precisely the characteristic signature of self-absorbed synchrotron emission, which is typically associated with the presence of a compact jet \citep{marscher1985}. In this scenario, the continuum is optically thick at low frequencies and the flux density distribution is constant for the ideal case ($\alpha \sim 0$), although a non-flat trend and ripples can be seen in this range due to optical depth effects and inhomogeneities in the jet structure \citep{band1986,kool1989,vuillaume2018}. There is a break in the continuum distribution at a certain turnover frequency when the synchrotron emission becomes transparent, leading to a decreasing flux with increasing frequency. In this domain, the slope of the power-law continuum is directly related with the energy distribution of the accelerated particles and their cooling processes \citep{blandford1979}. NGC\,1052 in the middle right panel of Fig.\,\ref{fig_seds} is one of the best examples supporting this scenario, since the lack of dust in the centre allows us to directly trace the flat-to-inverted continuum shape of the nuclear emission up to the far-UV range. Additionally, the characteristic turnover knee at $\gtrsim 20\, \rm{\micron}$ is clearly revealed even in the large aperture measurements, due to the strength of the mid-IR nuclear emission relative to the host galaxy. Finally, the inverse Compton upscattering of synchrotron photons by the accelerated particles in the jet produces an additional spectral component at high energies known as synchrotron self-Compton.

An additional argument supporting the compact jet scenario in LLAGN is provided by the observed properties in polarised light. The nuclear emission is expected to show a relatively low degree of linear polarisation in the optically thick part of the spectrum, oriented with an angle parallel to the direction of the jet axis, while the optically thin continuum at higher frequencies should have a larger degree of linear polarisation oriented along the direction perpendicular to the jet \citep{pacholczyk1970,blandford1979,marscher2014}. This has been demonstrated for the cases of M87 and NGC\,1052. The core of M87, observed with VLBA at $\sim 0.4\, \rm{mas}$ resolution, shows a linear polarisation degree of $0.2$--$0.6$\% at $43\, \rm{GHz}$ oriented along $66$--$92^\circ$ \citep{park2021}, in contrast with the $2$--$10$\% degree measured at $221\, \rm{GHz}$ ($\sim 1''$ resolution) and in the optical range ($\sim 0\farcs1$--$0\farcs6$), with $\sim 0$ to $-8^\circ$ orientation \citep{goddi2021,avachat2016,fresco2020}, albeit with some variability in the optical position angle \citep{avachat2016}. The same scenario applies to NGC\,1052, showing $< 0.2$\% at $221\, \rm{GHz}$ in the optically thick continuum at $\sim 1''$ resolution \citep{goddi2021} and $\sim 4$\% with a position angle of $120^\circ$ at $\sim 10\, \rm{\micron}$ in the optically thin range at $\sim 0\farcs5$ resolution \citep{jafo2019}.

This framework is commonly adopted to analyse the SED of blazars \citep{maraschi1992,sikora1997} and radio galaxies \citep{konopelko2003,petropoulou2014}, as well as that of X-ray binaries \citep{zdziarski2001,veledina2011}, but it also describes remarkably well the shape of the sub-arcsecond flux density distribution for the 8 LLAGN in Fig.\,\ref{fig_seds}. The self-absorbed synchrotron plus synchrotron self-Compton radiation has been represented in this Figure by a broken power law distribution with three components (dashed-black line), one intended for the radio continuum, a second one for the IR-to-optical/UV range, and a third one for the X-rays. Although the spectrum of inverse Compton radiation flattens rapidly close to the average energy of upscattered photons \citep{condon2016}, we adopted a simpler power-law distribution since the frequency sampling in our case does not appear to reach this domain (see Section\,\ref{fledgling} for a more detailed modelling of the X-ray emission). The piecewise power-law function has been fitted to the sub-arcsecond flux measurements of these nuclei using 6 free parameters, that is two power-law spectral indices, plus two frequencies at the knees of the broken power law and their associated fluxes. The index in the IR-to-optical/UV range is determined once the other parameters have been fixed. A bootstrap method was used to determine the parameter uncertainties by performing 1000 fits on synthetic flux density distributions obtained from Gaussian-based deviations of the measured values based on the observational errors (grey-shaded area in Fig.\,\ref{fig_seds}). The best-fit values obtained for the broken power-law distribution are shown in Table\,\ref{tab_fit}. For the case of the Sombrero galaxy, we excluded the optical and near-IR flux values from the fit to avoid the thermal bump, however a more elaborated model including the disc contribution will be discussed later in Section\,\ref{disc}.

When compared to the \textit{continuum spectra} of radio galaxies, two major differences arise in the case of LLAGN: \textit{i)} the IR-to-optical/UV continuum is significantly steeper with a slope index of $-3.7 \lesssim \alpha_0 \lesssim -1.3$, in contrast with the $\alpha_0 \sim -0.7$ index expected for standard particle acceleration scenarios \citep{drury1983}, which applies to the optically thin synchrotron continuum in radio galaxies \citep{eckart1986}; \textit{ii)} the turnover frequency above which the optically thick and roughly flat radio continuum becomes transparent and bends to the inverted optically thin continuum. While spectral turnovers in radio galaxies and compact radio cores usually happen at frequencies below $\sim 10\, \rm{GHz}$ \citep{eckart1986,dezotti2010,orienti2016,odea2021}, we detect this transition at $\sim 10$--$30\, \rm{\micron}$ ($\sim 10$--$30\, \rm{THz}$; see Table\,\ref{tab_fit}). Nevertheless, similar values have been reported for the core of radio-loud AGN \citep[e.g.][]{meisenheimer2007,migliori2011,grandi2016}. This is indicative of a very compact size for the synchrotron-emitting region \citep{dallacasa2000,tinti2005}. We note that, in the case of blazars, the high turnover frequencies observed are likely overestimated due to beaming effects. The lowest turnover frequencies are found for the Sombrero galaxy and M87 at $2\, \rm{THz}$ and $200\, \rm{GHz}$, respectively, which are precisely the two objects with the lowest Eddington rates in our sample ($\lambda_{\rm Edd} < -5$). These are also two of the three nuclei with the flatter IR-to-optical/UV continuum, $\alpha_0 \sim -1.7$ and $-1.3$, respectively.

\subsection{Beyond the far-UV continuum}\label{cloudy}
Fine-structure lines in AGN are originated by collisionally excited gas in the narrow-line region, since the low electron densities ($n_{\rm e} \sim 10^{2-5}\, \rm{cm^{-3}}$) permit the radiative decay of these transitions, unlike the high-density gas in the broad-line region. The nebular lines intensities respond mainly to the shape and strength of the ionising continuum and the physical conditions of the emitting gas, and therefore they are widely used to build diagnostics and determine gas properties such as temperature and density \citep{osterbrock2006}. Of special interest for this project are the nebular lines in the IR range which, in contrast with optical lines, are essentially unaffected by dust extinction --\,typically present in the centres of galaxies\,-- and their emissivities are weakly dependent on the gas temperature \citep{spinoglio1992}. The latter is due to the proximity of the ground state to the energy levels involved in these transitions, which become populated even for warm plasmas ($T_{\rm e} \gtrsim 1000\, \rm{K}$). An additional advantage is the wide range in the ionisation potentials of the ionic species that produce these lines, from a few $\rm{eV}$ to $> 100\, \rm{eV}$ in the case of coronal lines, which makes them ideal tracers of the black hole accretion \citep[e.g.][]{melendez2008,spinoglio2021,mordini2021}.

Direct measurements of the ionising continuum in the far-UV range are challenging due to the light extinction by dust and the intrinsic faintness of LLAGN at these wavelengths. Shortwards of $912$\,\AA, the atomic hydrogen located along the line of sight causes a heavy absorption in the nuclear ionising continuum, which becomes completely extinguished. However, partial information on the shape of this continuum remains encoded in the intensities of the nebular lines. These are associated with different ionic species whose ionisation potentials fall within the EUV range. Thus, diagnostics based on the nebular line ratios can be used to infer the shape of the LLAGN continuum beyond the range than can be directly observed. This is particularly useful for galaxies where the optical/UV nucleus is obscured by dust in the centre or by dust lanes in the host galaxy blocking the line of sight to the nucleus (e.g. NGC\,3169; see also \citealt{prieto2014,prieto2021}, \citealt{mezcua2016}). In particular, the unique characteristics of nebular lines in the IR have been exploited in the past to sense the shape of the ionising continuum in AGN \citep{alexander1999,alexander2000,melendez2011,jafo2021}.
\begin{figure*}
  \centering
  \includegraphics[width = 0.5\textwidth]{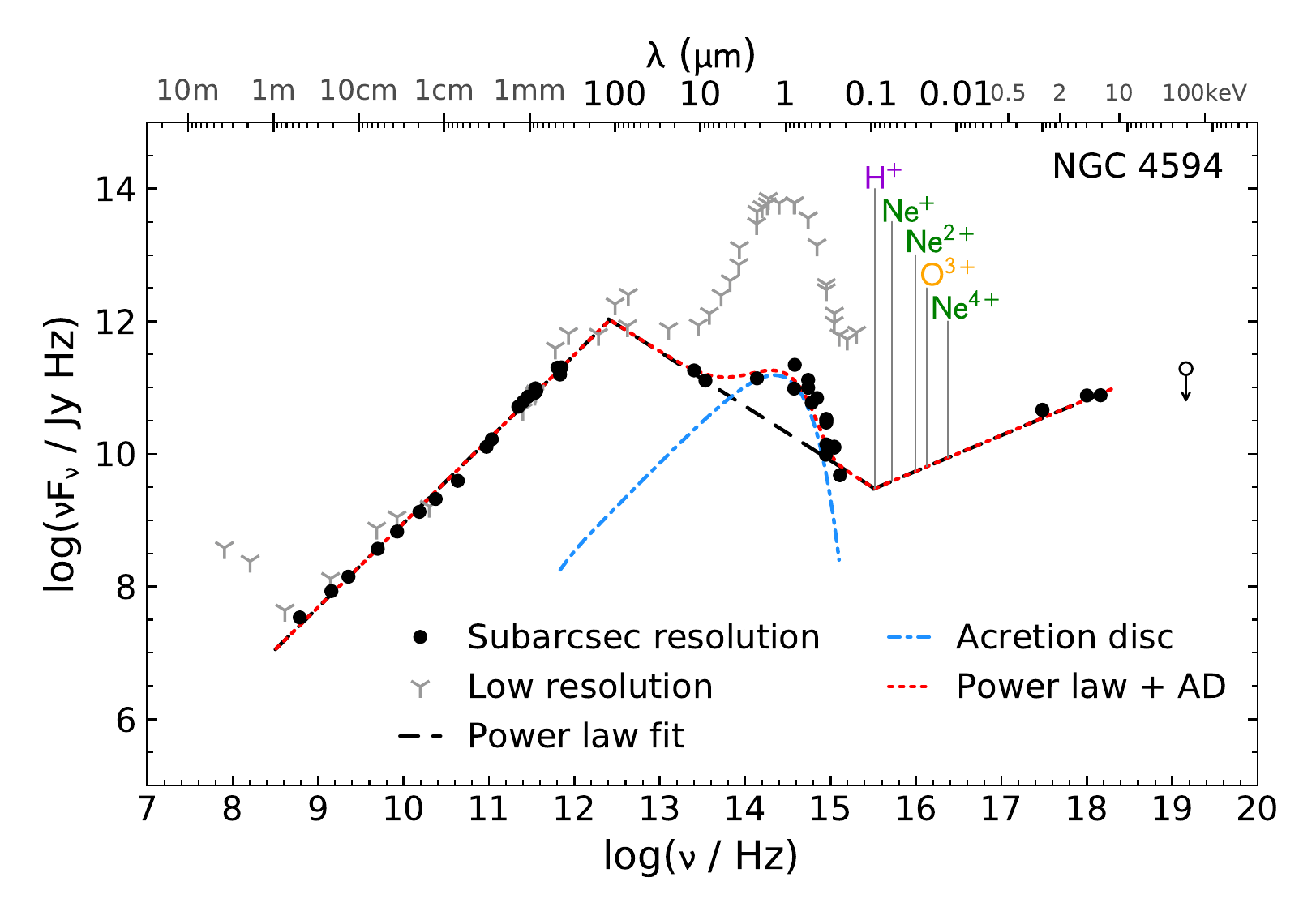}~
  \includegraphics[width = 0.5\textwidth]{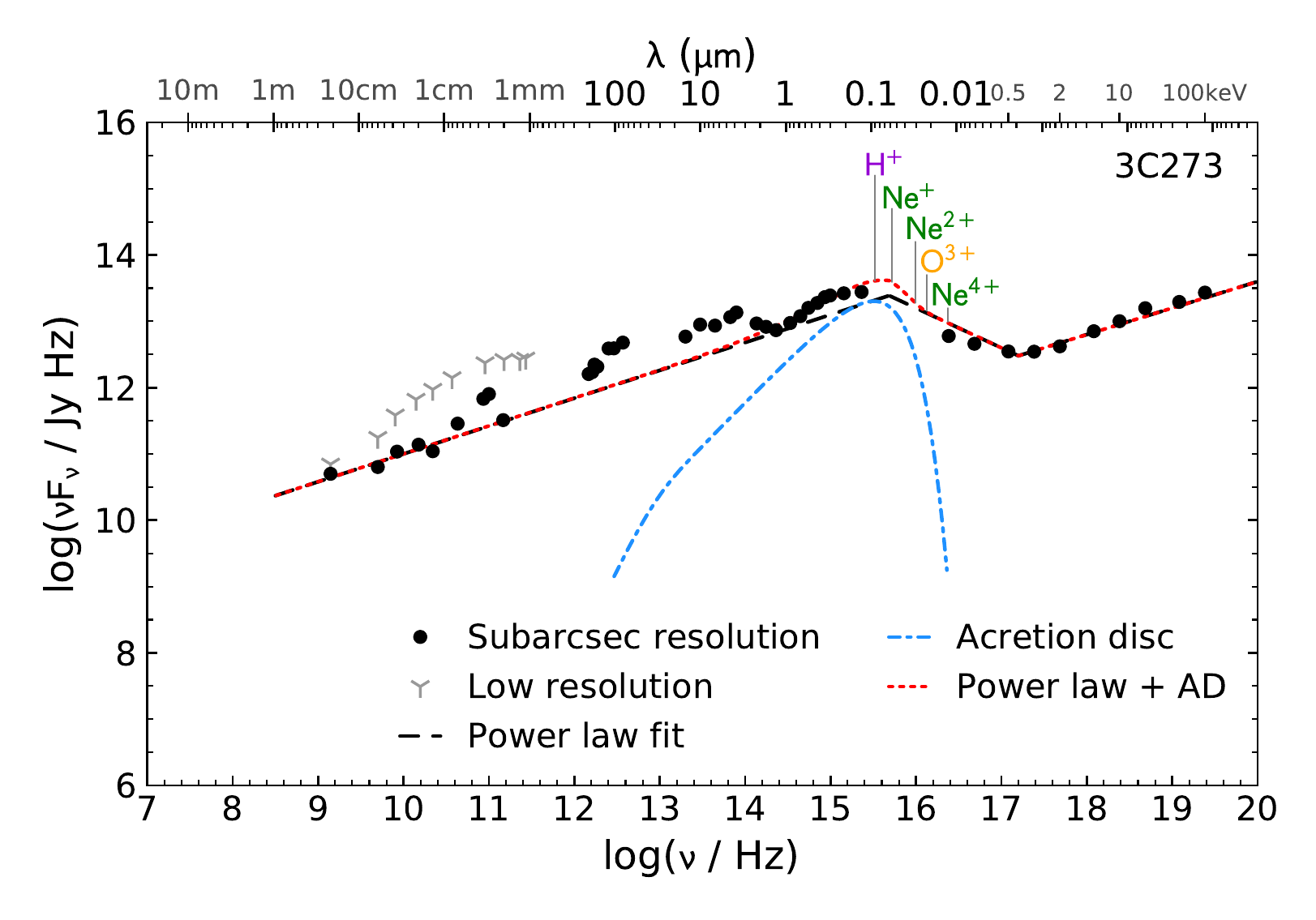}
  \caption{Best-fit to the high resolution data of NGC\,4594 (\textit{left}), consisting on a broken power-law (dashed-black line) plus a simulated accretion disc (dash-dotted blue line). The sum of both contributions (dashed-red line) is used as input ionising radiation field for the \textsc{Cloudy} simulations. Vertical coloured lines mark the ionisation energy of the ionic species used in the diagnostics diagrams. The disc assumes an inner (truncation) radius of $10\, \rm{r_g}$, and an accretion rate of $\dot{m} = 3 \times 10^{-5}\, \dot{\rm m}_{\rm Edd}$ for a BH mass $M_{\rm BH} = 10^{9}\, \rm{M_\odot}$. A similar analysis is done for the quasar 3C\,273 (\textit{right}). Disc parameters: inner radius = $6\, \rm{r_g}$, $\dot{m} = 0.2\, \dot{\rm m}_{\rm Edd}$ and $M_{\rm BH} = 7 \times 10^{8}\, \rm{M_\odot}$.}\label{fig_n4594}
\end{figure*}

In this work we have built diagnostic diagrams based on the observed ratios of mid-IR nebular lines detected by \textit{Spitzer}/IRS. These are then compared with the ratios predicted by photo-ionisation simulations computed with \textsc{Cloudy} \citep{ferland2017}. The novelty in our approach when compared with previous attempts to model the spectra of LLAGN is to use, as input ionising continuum, a broken power law derived from the fits to the high-angular resolution spectra of LLAGN, that is the dashed-line piecewise power-law distributions shown in Fig.\,\ref{fig_seds}. In this scenario, the ionising continuum driving the excitation of the nebular gas would be a combination of a steep power-law continuum in the optical, extrapolated beyond the far-UV range, and the rising synchrotron self-Compton continuum at higher frequencies. This approximation seems reasonable at first order from the trend suggested by the IR-to-UV continuum and the X-ray measurements, as shown in Fig.\,\ref{fig_n4594} for the case of the Sombrero galaxy.

As a first step, the best-fit broken power law continuum obtained for Sombrero is adopted as reference template for the ionising continuum in \textsc{cloudy} photo-ionisation simulations (dashed-black line in Fig.\,\ref{fig_n4594}, left panel; Table\,\ref{tab_fit}). This is characterised by a spectral slope of $\alpha_0 \sim -1.7$ up to $1.9 \times 10^{15}\, \rm{Hz}$ and $\alpha_{\rm x} \sim -0.7$ at higher frequencies. The fit relies on the high angular resolution dataset ($\lesssim 0\farcs4 = 18\, \rm{pc}$ at the distance of NGC\,4594) compiled for this work (black dots in Fig.\,\ref{fig_n4594}). The choice of Sombrero as a reference for low-excitation nuclei has a threefold motivation: \textit{i)} the steep optical continuum found to be common among the LINER class (of which NGC\,4594 is a prototypical example; \citealt{fabbiano1997,nicholson1998,hada2013}), and the inferred slope of the ionising continuum ($\alpha_{\rm x} \sim -0.7$), which is representative of the average slope in our sample (Table\,\ref{tab_fit}); \textit{ii)} the superior sampling in wavelength achieved for this galaxy due to the lack of dust obscuration, which allows us to detect the near-UV continuum; and \textit{iii)} the detection of a thermal bump component peaking at $\sim 1\, \rm{\micron}$, which allows us to investigate the possible contribution of the cold disc emission to the gas excitation. Although we show in this work the model predictions for the case of NGC\,4594, the simulations computed using the parameters in Table\,\ref{tab_fit} for the rest of LLAGN in the sample are in agreement. To quantify the variation in the line ratios predicted when different sources templates are used for the simulations, we produced \textsc{cloudy} models for the nuclei with the largest difference in the spectral index of the ionising continuum, that is M87 ($\alpha_{\rm x} = -1.2$) and NGC\,1052 ($\alpha_{\rm x} = -0.44$). The line ratios predicted for different sources show, on average, small differences ($\lesssim 0.2\, \rm{dex}$), suggesting that the lack of a prominent accretion disc in their continuum limits the capability of these nuclei to ionise the narrow-line region gas.

To investigate the effect of a cold accretion disc in the gas excitation, we produced a second set of photo-ionisation models for NGC\,4594 including the contribution of a disc assuming the Shakura-Sunyaev formalism (Fig.\,\ref{fig_n4594}). The thermal bump in the sub-arcsecond resolution SED of Sombrero has been fitted using a BH mass of $M_{\rm BH} = 7 \times 10^{8}\, \rm{M_\odot}$ \citep{jardel2011}, with an inner (truncation) radius of $10\, \rm{r_g}$, and a mass accretion rate of $\dot{\rm m} = 3 \times 10^{-5}\, \dot{\rm m}_{\rm Edd} = 4.4 \times 10^{-4}\, \rm{M_\odot\,yr^{-1}}$. These values are in agreement with the results obtained by \citet{pellegrini2003} and \citet{satyapal2005} based on the analysis of the X-ray spectrum. The truncated accretion disc is then added to the underlying power-law continuum used as input for \textsc{Cloudy} simulations. Finally, a similar procedure was performed for the bright quasar 3C\,273. This has been adopted as a template to model the photo-ionisation produced by the big blue bump continuum, which is prominent in the optical/UV range for this quasar nucleus (Fig.\,\ref{fig_n4594}). To complete the far- to EUV gap in the flux density distribution provided by \citet{prieto2010}, we followed a similar approach as in NGC\,4594. In the case of 3C\,273, the continuum emission has been modelled using a piecewise power-law distribution plus a Shakura-Sunyaev accretion disc (Fig.\,\ref{fig_n4594}, right), adopting a $M_{\rm BH} = 7 \times 10^{8}\, \rm{M_\odot}$ \citep{bentz2015}. From the best-fit model to the SED we obtain a disc inner radius of $6\, \rm{r_g}$ and a mass accretion rate of $\dot{\rm m} = 0.2\, \dot{\rm m}_{\rm Edd} = 3\, \rm{M_\odot\,yr^{-1}}$ \citep{paltani2005}.

To simulate the physical conditions of the gas exciting the IR fine-structure lines, we produced several grids of photo-ionisation models sampling a range in ionisation parameter ($-1.5 < \log U < -3$) and density ($2 < \log(n_{\rm e}/\rm{cm^{-3}}) < 4$) for the LLAGN template based on NGC\,4594 and the quasar template inspired by 3C\,273. Plane-parallel geometry and solar abundances \citep{asplund2009} are assumed, and Polycyclic Aromatic Hydrocarbons (PAHs) are also included. All models have been run with constant pressure, which forces the density to be variable within the nebula with the distance to the radiation source. The cosmic microwave background is also included as part of the incident radiation, as well as the contribution from a cosmic ray background similar to that in the Milky Way. As a stopping criterion we used hydrogen column density $N_{H} = 10^{23}\, \rm{cm^{-2}}$. Calculations have been performed with the \textsc{Cloudy} version C17.01, last described in \cite{ferland2017}, assisted by the py\textsc{Cloudy} library \citep{morisset2013}.

The validity of this hypothesis is assessed in Sections\,\ref{hidden} and \ref{disc} by comparing the observed line ratios with the predicted values from \textsc{Cloudy} simulations. Since the EUV continuum cannot be directly probed due to heavy absorption from hydrogen, we rely on this indirect method to infer the spectral shape of LLAGN beyond the high-angular resolution measurements in the far-UV.

We note that additional contributions to the gas excitation due to star formation or shocks are not included in our models. Bright star-forming regions are not detected in the adaptive optics images of the central few arcsec of these galaxies, which corresponds to the \textit{Spitzer} slit aperture, except for the case of NGC\,1097 longwards of $\sim 19\, \rm{\micron}$ \citep[see fig.\,1 in][]{bernard-salas2009}. In most cases the underlying stellar population is old and does not contribute significantly to the ionising radiation \citep[e.g.][for the case of NGC\,1052]{jafo2011b}. On the other hand, the possible contribution from shocks is discussed in Section\,\ref{shocks}, where we compare our predictions with the models from \citet{contini2001}. Nevertheless, the mid-IR line profiles in the \textit{Spitzer} spectra do not show significant shifts in velocity or broad profiles, suggesting that the contribution from shocks to the gas excitation within the innermost $\lesssim 10''$, if present, is limited to relatively low velocities ($\lesssim 500\, \rm{km\,s^{-1}}$). Finally, the extended radio emission from the kpc-scale lobes detected in some of the galaxies (e.g. NGC\,1052 and M87) has not been taken into account in the photo-ionisation models, since the continuum radiation at these wavelengths does not contribute to the ionisation. Nevertheless, a small contribution to the non-thermal UV continuum in the case of M87 could be provided by the hotspots detected in the inner jet of this galaxy.
%Although shocks might be important for certain objects (e.g. the case of NGC\,1386 in \citealt{rodriguez-ardila2017}; see Section\,\ref{shocks}), our approach in this work is to provide a simpler but more general description of the ionising continuum in LLAGN, rather than developing sophisticated models and detailed analyses of individual sources \citep[e.g.][]{dopita2015}. 
 
\begin{figure*}
\centering
\includegraphics[width = 0.5\textwidth]{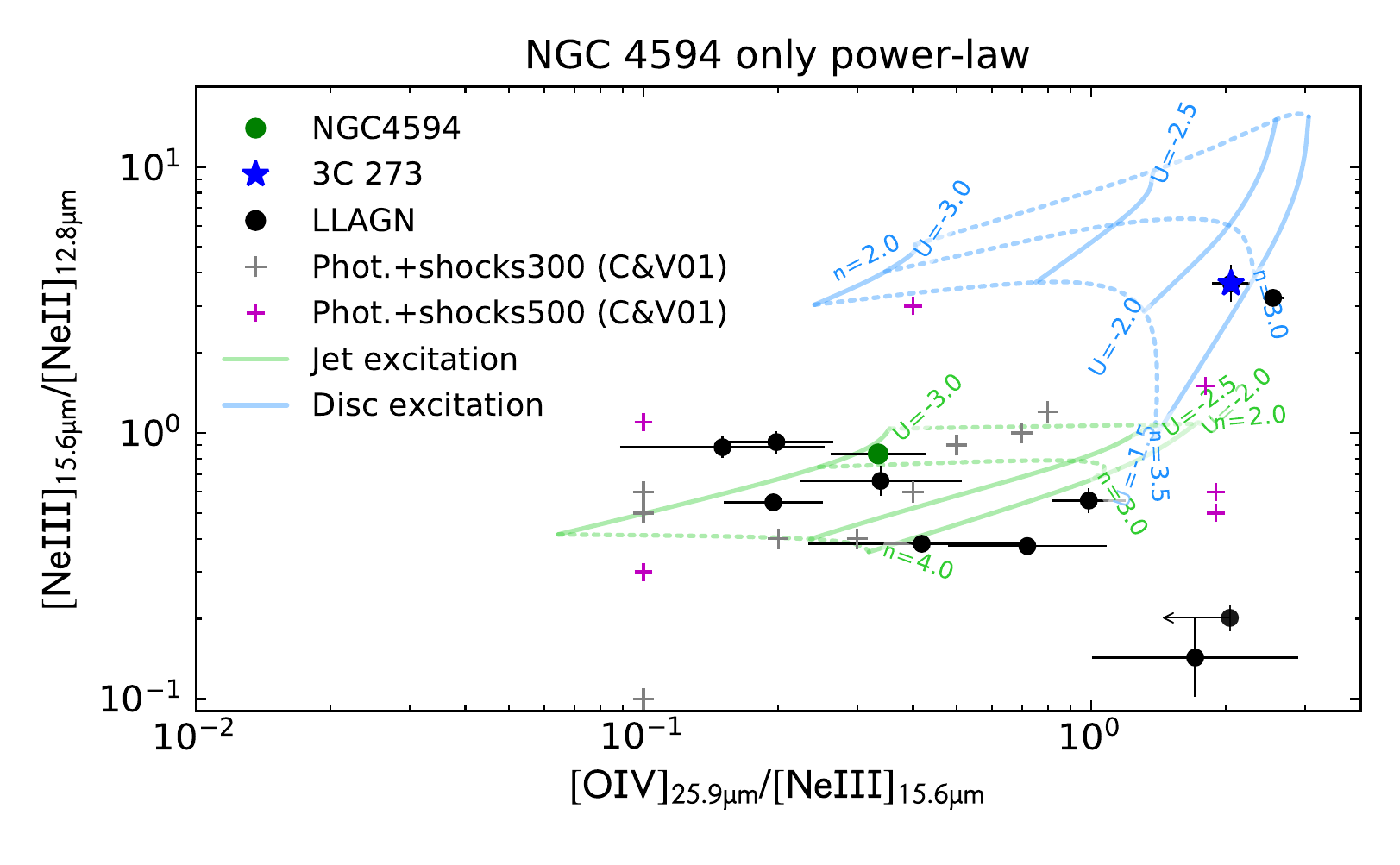}~
\includegraphics[width = 0.5\textwidth]{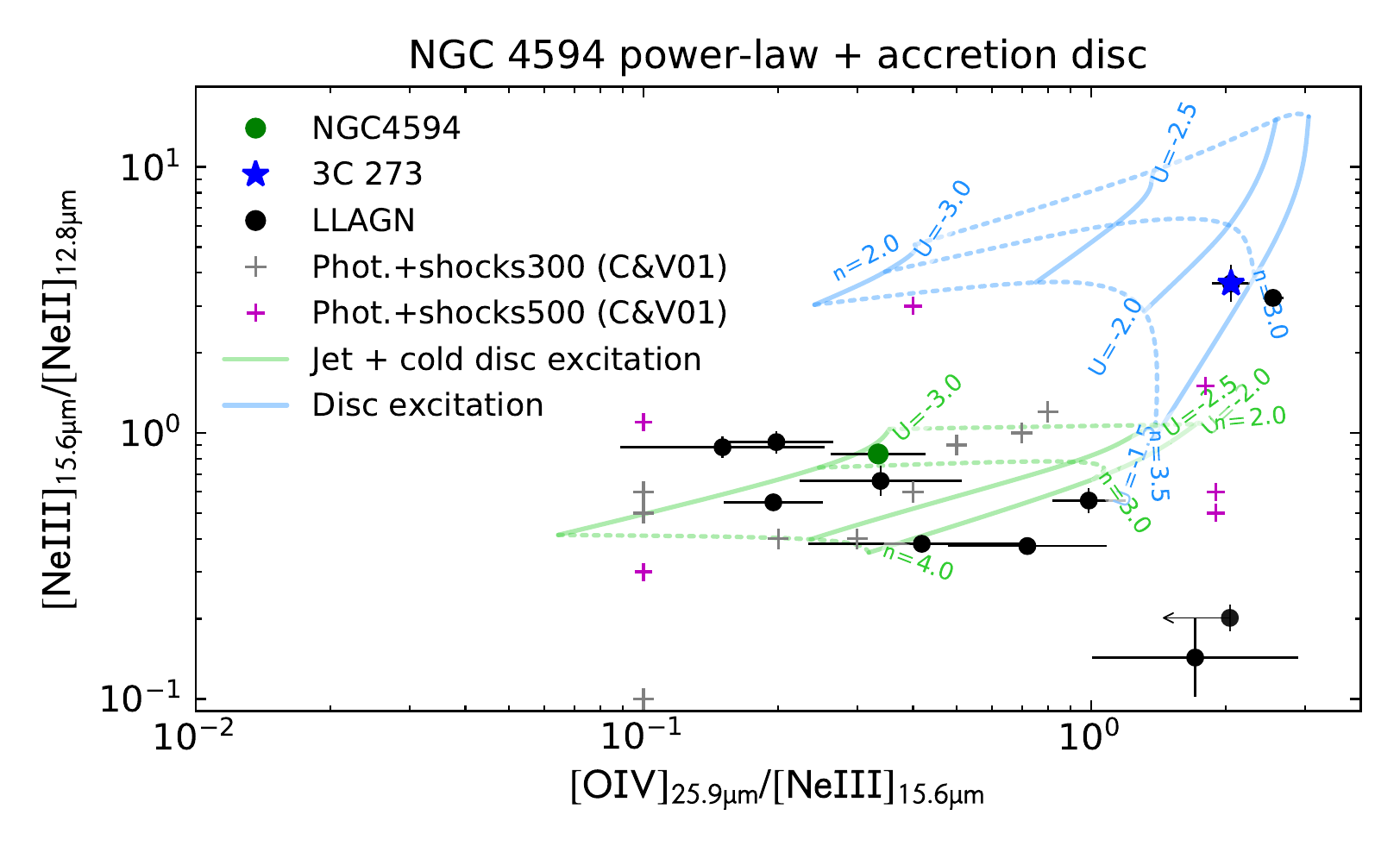}
\caption{Diagnostic diagrams for the LLAGN sample with photo-ionisation models of the Sombrero galaxy overlaid. Black dots show the observed flux ratios with their associated uncertainty. The green grid is constructed by interpolation of the values of certain line ratios which correspond to different values of ionisation parameter U and hydrogen density $n_0$. These ratios are calculated with \textsc{Cloudy} simulations. In the left panel, the input radiation field for these models is the power law fit to NGC\,4594 shown in Fig.\,\ref{fig_n4594}, while in the right panel the accretion disc (also shown in Fig.\,\ref{fig_n4594}) is included. The blue grid represents the model predictions using the 3C\,273 power-law fit (Fig.\,\ref{fig_n4594}) as input radiation source. Predictions for models including both photo-ionisation and shocks, from \citet{contini2001}, are shown by the crosses in grey (shock velocity of $v = 300\, \rm{km\,s^{-1}}$ and pre-shock density of $n_0 = 300\, \rm{cm^{-3}}$), and purple ($v = 500\, \rm{km\,s^{-1}}$) colour. %, orange ($v = 1000\, \rm{km\,s^{-1}}$), and cyan ($v = 1500\, \rm{km\,s^{-1}}$) colour.
The various models correspond to different cloud sizes and radiation field intensities (see Section\,\ref{shocks}).}\label{fig_diag}
\end{figure*}

\section{Discussion}\label{discuss}

The shape of the nuclear continua in the nearest LLAGN, shown in Fig.\,\ref{fig_seds}, departs from the classical AGN templates derived for quasars and Seyfert galaxies, characterised by a blue bump, the signature of the accretion disc, and an IR bump due to dust-reprocessed emission from the disc \citep[e.g.][]{prieto2010}. The continuum spectra of LLAGN do not show these thermal bumps, high-spatial resolution data ($\sim 10\, \rm{pc}$) reveal instead a rather featureless continuum compatible with a broken power law over more than ten orders of magnitude in frequency across the electromagnetic spectrum. This characteristic shape is indicative of self-absorbed synchrotron emission from a compact jet in the IR-to-optical/UV and its associated inverse Compton emission in the X-rays. This result is in line with earlier results from \citet{ho1996a} and \citet{ho2008} suggesting that the optical/UV spectrum of LLAGN is compatible with a power law, and especially the more detailed theoretical modelling of prototypical objects of this class such as M81 \citep{markoff2008} and M87 \citep{prieto2016}. We further find the slope of the IR-tooptical/UV spectrum to be considerably steeper than the classical slope produced by standard synchrotron jet cooling. Presumably, such a step spectrum is related with the lack of accelerated particles at the base of the jet (see Section\,\ref{fledgling}). The thermal contribution from the disc is absent in most cases. The disc is detected peaking in the near-IR for the case of Sombrero (Fig.\,\ref{fig_n4594}), suggesting that this structure might be cold and possibly truncated in most LLAGN, as expected from ADAF theory \citep{narayan1994}. Dust emission appears to have a minor contribution as well, because a thermal component seems to be missing in the nuclear mid-IR emission. This has been proved for the case of NGC\,1052 using the highest subparsec resolution attainable with mid-IR interferometry \citep{jafo2019}.

In this Section we discuss the shape of the ionising continuum in LLAGN as seen by diagnostics based on the IR line ratios (Section\,\ref{hidden}). A comparison with previous works using photo-ionisation plus shock models is addressed in Section\,\ref{shocks}. The contribution of the accretion disc to the gas excitation in LLAGN is investigated in Section\,\ref{disc}, where we discuss also previous works supporting the existence of Seyfert-like discs in LLAGN. Finally, a more physical description of the LLAGN continuum emission across the electromagnetic spectrum is proposed in Section\,\ref{fledgling}.

\subsection{The hidden ionising continuum}\label{hidden}
Direct observations do not allow us to probe the shape of the ionising continuum in LLAGN, mainly due to hydrogen absorption. A first guess suggests that the steep IR-to-optical/UV continuum seen in Fig.\,\ref{fig_seds} would flatten at higher frequencies when the rising inverse Compton emission overcomes the decreasing synchrotron contribution, as has been suggested by \citet{jafo2021}. However, a more elaborated analysis is still required in order to discard a hypothetical thermal contribution from a blue bump to the unseen EUV continuum or perhaps from the soft X-ray excess component \citep{singh1985,arnaud1985}. This analysis can be done by comparing the observed line ratios with the predictions from the photo-ionisation models described in Section\,\ref{cloudy}.

The left panel of Fig.\,\ref{fig_diag} shows the observed IR nebular line ratios of [\ion{Ne}{iii}]$_{15.6}$/[\ion{Ne}{ii}]$_{12.8}$ versus [\ion{O}{iv}]$_{25.9}$/[\ion{Ne}{iii}]$_{15.6}$ for the nine LLAGN included in our sample (black circles). These line ratios are optimal to probe the shape of the EUV continuum because the ionic species involved respond to the continuum emission in the $\sim 22$--$77\, \rm{eV}$ range. The green grid in this figure represents the predicted line ratios for the models using the non-thermal continuum in NGC\,4594 (dashed-black line in Fig.\,\ref{fig_n4594}, left) as input spectrum, assuming a certain range in density (dotted-green lines) and ionisation parameter values (solid-green lines). Our aim with these models is to probe the line ratios expected for the ionisation produced by a compact jet, that is, a pure synchrotron plus inverse Compton continuum without the thermal contribution from the accretion disc. The jet excitation models in the left panel of Fig.\,\ref{fig_diag} show [\ion{Ne}{iii}]$_{15.6}$/[\ion{Ne}{ii}]$_{12.8} \sim 0.4$--$1$ and [\ion{O}{iv}]$_{25.9}$/[\ion{Ne}{iii}]$_{15.6} \sim 0.07$--$1.5$, in agreement with the overall distribution of the observed line ratios. In particular, the contribution from the inverse Compton continuum in the EUV range is key to reproduce the behaviour of these lines, since this component likely dominates the ionising continuum in LLAGN. There are two objects that fall below the simulated grid in the [\ion{Ne}{iii}]$_{15.6}$/[\ion{Ne}{ii}]$_{12.8}$ axis. One source is NGC\,404, which remains undetected in [\ion{O}{iv}]$_{25.9}$ and its nuclear optical-to-IR continuum appears to be dominated by the host galaxy starlight (see Section\,\ref{shape}). The second case is NGC\,1097, in which the \textit{Spitzer} slit partially includes the circumnuclear starburst ring found in this galaxy \citep{bernard-salas2009}. This ring likely contaminates the [\ion{Ne}{ii}]$_{12.8}$ emission, explaining the location of this source in the vertical axis.

Additionally, the blue grid in Fig.\,\ref{fig_diag} represents the line ratios predicted for disc photo-ionisation, adopting the 3C\,273 template as input continuum (dashed-black line in Fig.\,\ref{fig_n4594}, right), sampling also a similar range in density (dotted-blue lines) and ionisation parameter (solid-blue lines). As expected, the observed line ratios for 3C\,273 are consistent with the model-based predictions using the ionising continuum inferred from its SED. NGC\,1386 is the only LLAGN that falls close to the range delimited by the disc photo-ionisation models. In this case, the remarkable intensities observed from high-excitation lines such as [\ion{O}{iv}]$_{25.9}$ and [\ion{Ne}{v}]$_{14.3,24.3}$ in the \textit{Spitzer} spectra (Fig.\,\ref{fig_mirspec}) are explained by the presence of strong shocks caused by the jet-ISM interaction found in this galaxy, which are required to explain, for instance, the detection of extended emission in the [\ion{Si}{VI}]$_{1.96}$ coronal line \citep{rodriguez-ardila2017}.

The analysis performed in this Section suggests that the ionising continuum defined by the extrapolation of both the synchrotron power-law and the inverse Compton components, shown in Fig.\,\ref{fig_n4594}, is able to reproduce the observed line ratios for the LLAGN in our sample. This implies that an excess in the EUV continuum above the adopted power-law continuum is unlikely, as it would overestimate the observed line ratios. We can also infer the lack of a prominent accretion disc in NGC\,3169, where the optical/UV continuum is obscured by dust and therefore cannot be directly observed. The line ratios for this source fall also on the \textsc{cloudy} grid that reproduces the values measured for the rest of the LLAGN sample, thus we conclude that the accretion disc is not relevant in this nucleus. For clarity, we only show in Fig.\,\ref{fig_diag} the photo-ionisation predictions for the models based on the template of NGC\,4594, nevertheless the results obtained adopting any of the other sources in Fig.\,\ref{fig_seds} as input continuum for the simulations are in agreement. For instance, the median difference between photo-ionisation predictions for M87 and NGC\,1052 --\,those with the largest difference between their ionising continuum slope $\alpha_{\rm x}$ in Table\,\ref{tab_fit}\,-- is of about $0.2\, \rm{dex}$ for the line ratios [\ion{Ne}{iii}]$_{15.6}$/[\ion{Ne}{ii}]$_{12.8}$ and [\ion{O}{iv}]$_{25.9}$/[\ion{Ne}{iii}]$_{15.6}$. In all cases, most of the observed line ratios in LLAGN are covered by the model grids within the $-3 < \log U < -2$ and $2 < \log(n_{\rm e}/\rm{cm^{-3}}) < 4$ ranges.

%On a wider context, it is worth mentioning that LLAGN in our sample show an average spectral slope of $\alpha_{\rm x} = -0.7$ with a standard deviation of $0.2$ in the $\nu \sim 10^{15}$--$10^{18}\, \rm{Hz}$ range ($\sim 3000$\,\AA \ to $10\, \rm{keV}$). This value is similar to the X-ray spectral slope of the Compton thick AGN population that dominates the cosmic X-ray background ($\alpha_{\rm x} \sim -0.9$; \citealt{ueda2003}, \citealt{gilli2007}, \citealt{barchiesi2021}). LLAGN represent about 1/3 of all galaxies in the local Universe \citep{ho2008}, if their number density is also high at high-z they may represent an important source of X-ray photons to the cosmic background besides the contribution of more luminous and obscured AGN.

\subsection{Photo-ionisation and shocks}\label{shocks}
In the previous Section we investigated the excitation of nebular IR lines as caused exclusively by photo-ionisation radiation from a broken power-law continuum, inspired by the continuum shape observed at high-spatial resolution in LLAGN shown in Fig.\,\ref{fig_seds}. Nevertheless, the presence of shocks as an important source of excitation in LLAGN has been proposed since their discovery in the early 1980's \citep{heckman1980,dopita1995,sugai2000,dopita2015,molina2018}.

In Fig.\,\ref{fig_diag} we compare our results with those predicted by \citet{contini2001} using the \textsc{suma} code \citep{viegas1994}, which provides a consistent approach to combine the effect of photo-ionisation radiation and a shock front simultaneously, rather than adding their separate contributions to the line ratios. These models are computed sampling a range in the values of shock velocities, pre-shock densities, geometrical thickness of the clouds, and intensity of the radiation field. When compared with the results shown in Section\,\ref{hidden}, the observed line ratios in LLAGN are in agreement with the \textsc{suma} predictions including mid- to low-velocity shocks ($\lesssim 300\,\rm{km\,s^{-1}}$; grey crosses in Fig.\,\ref{fig_diag}) for models with different radiation field intensities and nebular cloud sizes. We note, however, that the apparent sequence shown my the grey crosses in Fig.\,\ref{fig_diag} does not respond to a progressive increase in any of the model parameters, for example, the radiation field intensity. On the contrary, small changes in one of the model parameter values can result in a large variation of the predicted line ratios, due to the complex ionisation structure created in the simulated nebula when photo-ionisation and shocks are consistently combined \citep{contini2001}. In fact, some of the models (not shown in Fig.\,\ref{fig_diag}) predicted ratios falling far from both the LLAGN and the Seyfert excitation domains. On the other hand, some of the models with higher velocity shocks ($v > 500\, \rm{km\,s^{-1}}$) would be in agreement with the LLAGN ratios, although in these cases the velocity shift and/or the line broadening expected would have been resolved in the \textit{Spitzer} high-resolution spectra ($R = 600$, $\sim 500\, \rm{km\,s^{-1}}$). This is the case of NGC\,1386, where such high velocities in the coronal [\ion{Si}{vi}]$_{1.96}$ line are measured in a nuclear outflow that has been spatially resolved by SINFONI \citep{rodriguez-ardila2017}. The high excitation ratios in this nucleus seen in Fig.\,\ref{fig_diag} are likely ascribed to the shocks formed in the nuclear outflow of this galaxy. %For a more detailed modelling in the case of NGC\,1052 using multiple shock components we refer to \citet{dopita2015}.

Overall, models combining photo-ionisation and shocks are also in agreement with the observed line ratios in LLAGN when relatively weak radiation fields and slow shocks are considered. In particular, the models computed by \citet{contini2001} adopt a Seyfert-like continuum spectrum for the radiation field, thus low-velocity shocks are required to boost the low-excitation lines. In contrast, our analysis suggests that a more realistic continuum distribution using a compact jet spectrum as input radiation source already predicts the observed line ratios in LLAGN. While shocks might still be relevant to explain the line ratios in sources with nuclear outflows, or perhaps to explain the ionising photon budget at a few tens of parsec distance from the nuclear source \citep{molina2018}, they might have a minor contribution to the line excitation in the innermost $\lesssim 10\, \rm{pc}$ for the majority of the LLAGN population. The arrival of the \textit{James Webb Space Telescope}, successfully launched in December 2021, will allow us to test this and other scenarios by probing diagnostics involving higher-ionisation species such as [\ion{Ne}{v}]$_{14.3,24.3}$ and possibly also the coronal lines in the near- to mid-IR range \citep[e.g.][]{prieto2022}, which are typically too faint to be detected by \textit{Spitzer} in LLAGN.

\subsection{A cold accretion disc}\label{disc}
From the studied sample, the only LLAGN that exhibits a clear thermal bump which can be associated with emission from the accretion disc is the Sombrero galaxy. The bump seen in the left panel of Fig.\,\ref{fig_n4594} peaks at $\sim 1\, \rm{\micron}$, and therefore would correspond to a cold ($\lesssim 3000\, \rm{K}$) and/or a truncated disc. In order to study the possible contribution of such a disc to the nebular line emission, we computed a second set of photo-ionisation simulation models adopting the dotted-red line the left panel of Fig.\,\ref{fig_n4594} as input ionising continuum (Section\,\ref{cloudy}). As can be seen by comparing the two models in Fig.\,\ref{fig_n4594} (green grids in the left and right panels), the photo-ionisation predictions are essentially identical in both cases, with a deviation between models of $< 0.9 \%$ and $< 0.4 \%$ in [\ion{O}{iv}]$_{25.9}$/[\ion{Ne}{iii}]$_{15.6}$ and [\ion{Ne}{iii}]$_{15.6}$/[\ion{Ne}{ii}]$_{12.8}$, respectively. This result should come as no surprise, because the observed thermal bump is too cold as for having any direct contribution to the total nebular emission or to alter significantly the ionisation structure of the nebula.

Our results are in contrast with those from \citet{maoz2007} and \citet{xu2011}, who propose the existence of a Seyfert-like accretion disc in LLAGN. This is based on the continuum flux ratios measured between UV and X-ray wavelengths, which are very similar to those of Seyfert 1 nuclei likely dominated by the accretion disc emission in this range. The excellent wavelength sampling of our high-angular resolution continuum spectra allows us to address this issue by comparing the spectral slope of the LLAGN in our sample with those obtained by \citet{maoz2007} and \citet{xu2011}. Following these works, we adopt the $\alpha_{\rm ox}$ index to quantify the UV/X-ray ratio, defined as:
\begin{equation}
  \alpha_{\rm ox} = \frac{\log[L_\nu (2500\, \text{\AA})/L_\nu (2\, \rm{keV})]}{\log[\nu(2500\, \text{\AA})/\nu(2\, \rm{keV})]}.
\end{equation}

\begin{figure}
  \centering
  \includegraphics[width = \columnwidth]{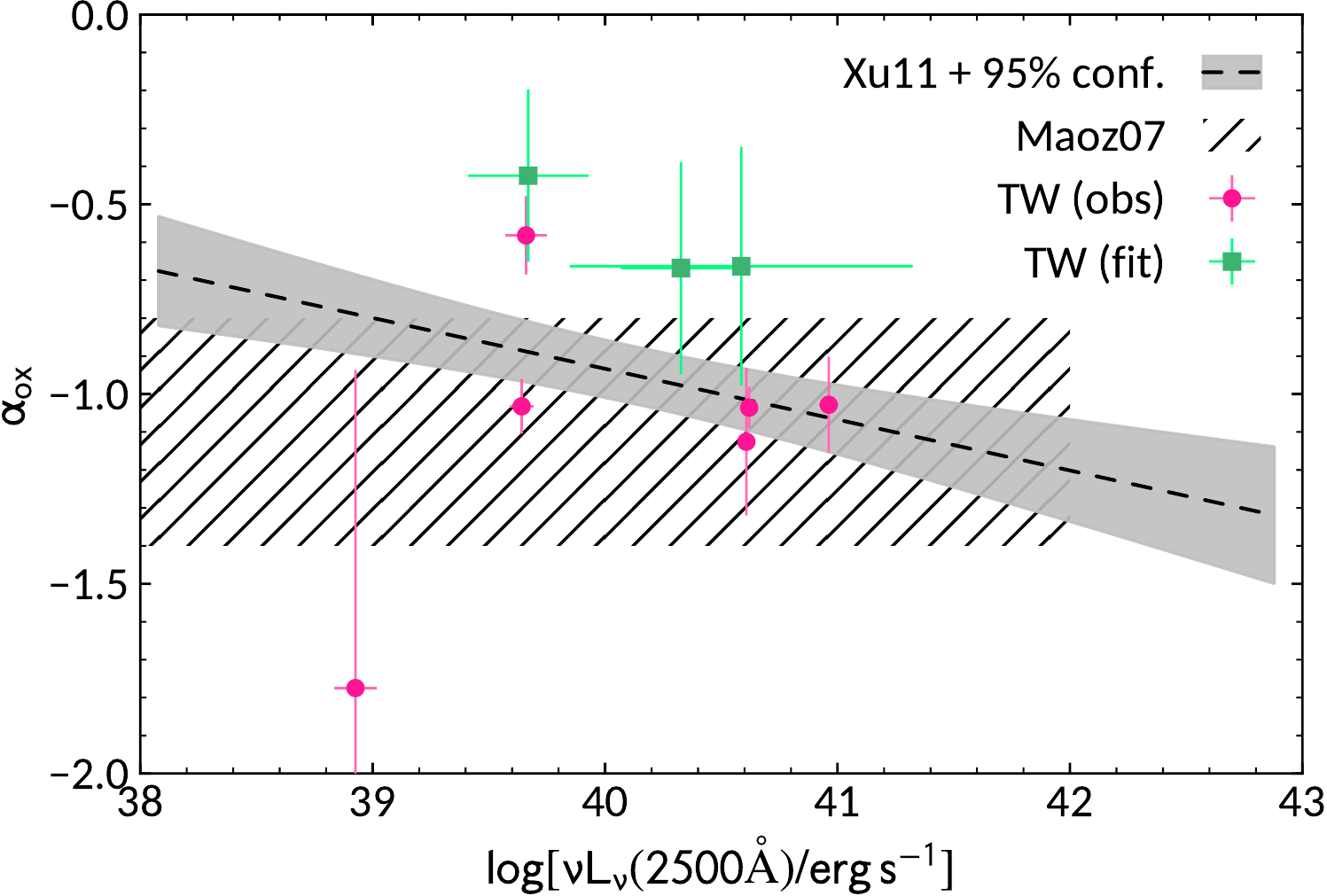}\vspace{0.3cm}
  \includegraphics[width = \columnwidth]{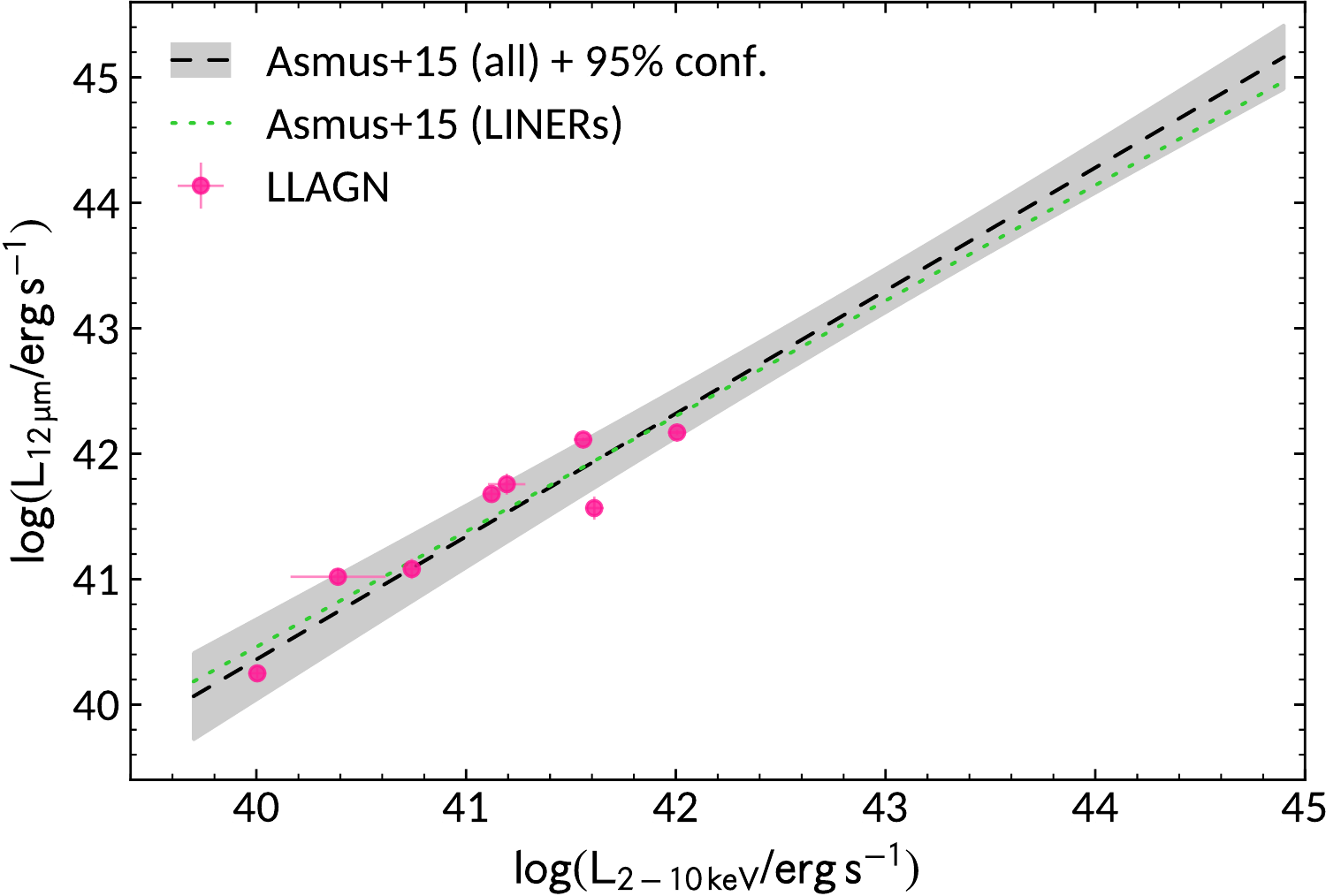}
  \caption{Optical-to-UV spectral slope in the $2500$\,\AA--$2\, \rm{keV}$ range ($\alpha_{\rm ox}$) versus the luminosity $\log(\nu L_\nu)$ at 2500\,\AA \ for the LLAGN in our sample (\textit{top panel}) using direct measurements (pink circles) and values inferred from the power-law fit (green squares). The majority of the sources are in agreement with the distribution of these parameters in the sample of \citet{maoz2007}, shown by the shaded-grey area, and the correlation derived by \citet{xu2011} for the Palomar sample. The comparison between the $12\, \rm{\micron}$ mid-IR and the $2$--$10\, \rm{keV}$ X-ray luminosities (\textit{bottom panel}) for the LLAGN in our sample (pink circles) fall within the 95\% confidence interval (shaded-grey area) of the correlation derived by \citet{asmus2015} for a sample of 152 nearby AGN (dashed-black line), including 19 LINER nuclei (dotted-green line for the correlation derived for the LINER sub-sample).}\label{fig_aox_IRX}
\end{figure}

The top panel in Fig.\,\ref{fig_aox_IRX} shows the distribution of the $\alpha_{\rm ox}$ values against the monochromatic luminosity at 2500\,\AA \ for our sample of LLAGN. Pink circles correspond to the $\alpha_{\rm ox}$ values derived from direct measurements, either from the broad-band fluxes or from the sub-arcsecond resolution \textit{HST} spectra shown in Fig.\,\ref{fig_seds}. For those objects where the UV range is not observed or detected (NGC\,1386, NGC\,3169, NGC\,4261), we used the extrapolation of the power-law fit to the IR-optical continuum (green squares). Although the real flux of these objects at 2500\,\AA \ is quite uncertain, the extrapolation allows us to identify what would be the location of these objects in the diagram of Fig.\,\ref{fig_aox_IRX} if the power-law continuum is representative of the nuclear flux at UV wavelengths. The shaded-grey area in the figure indicates the distribution of the $\alpha_{\rm ox}$ values measured by \citet{maoz2007} in his sample of 13 LLAGN ($-1.4 < \alpha_{\rm ox} < -0.8$ ), while the dashed-black line represents the fit derived by \citet{xu2011} for the correlation between the spectral slope and the UV luminosity for the sample of LLAGN in \citet{gu2009}, drawn from the Palomar sample \citep{ho1997} and the LINER catalogue of \citet{carrillo1999}. As shown by this figure, the majority of LLAGN in our sample are in agreement with the results by \citet{maoz2007} and \citet{xu2011}. However, the measured values are not driven by a thermal-like bump in the optical/UV range, but by a non-thermal continuum, represented by the emission of self-absorbed synchrotron plus inverse Compton in the optical/UV to the X-ray range (Fig.\,\ref{fig_seds}).

Additionally, we compared the X-ray to mid-IR luminosity ratio of our sources with the correlation derived by \citet{asmus2015} in the bottom panel of Fig.\,\ref{fig_aox_IRX}. This correlation, followed by Seyfert galaxies and quasars, suggests that the dust-reprocessed radiation from the AGN in the infrared might be tightly related with the accretion disc, who would also provide the seed photons that become upscattered at X-rays by the BH corona. Since LINERs seem to fit in the low-luminosity tail of this correlation, their central engines may also be similar to those of bright Seyfert galaxies. Again, the LLAGN in our sample are in excellent agreement with the X-ray to mid-IR correlation, however they do not show the disc feature in the UV, necessary to heat the dust, nor the IR bump that could be associated with the reprocessed dust emission (Fig.\,\ref{fig_seds}), although some dust contribution to the mid- and far-IR spectrum can still be detected in some of these nuclei mostly in the form of absorption or emission in the silicate dust features \citep[e.g.][]{mason2013}. For the cases of M87 and NGC\,1052 the non-thermal nature of the IR emission has been demonstrated by \citet{buson2009} and \citet{prieto2016}, and \citet{jafo2019}, respectively. On the other hand, NGC\,1097 and M87 do not show cold molecular gas emission in the innermost few parsecs \citep{izumi2017,simionescu2018}, expected if a torus would be present in these nuclei, albeit warm H$_2$ gas may be present in the innermost few parsecs \citep{mezcua2015}.

The reason why sources like M87 and NGC\,1052 still fit in the mid-IR to X-ray correlation is probably related with the energy balance associated with the particle cooling processes in these sources. As shown in Section\,\ref{fledgling}, for relatively low Lorentz factors in the accelerated electrons, of about $\gamma_{\rm e} \sim 500$, the mid-IR photons ($\sim 20\, \rm{\micron}$) upscattered by inverse Compton processes would fall within the X-ray range at $\sim 20\, \rm{keV}$, with a maximum cutoff around $60\, \rm{keV}$. This explains why a mid-IR to X-ray correlation is also present in compact jet dominated sources.

\subsection{An undeveloped jet}\label{fledgling}
The power-law piecewise distribution shown in Fig.\,\ref{fig_seds} provides a reasonable description, at first order, of the LLAGN continuum emission over $\sim 10$ orders of magnitude in wavelength, for the nuclei in our sample. This is further supported by the photo-ionisation models discussed in Section\,\ref{hidden}, which suggest that the hidden continuum shape should not be far from our power-law approximation used to fill the EUV gap unreachable by direct observations, discarding the presence of a hypothetical emission excess in this range associated with the accretion disc. The interpretation of the power laws is a self-absorbed synchrotron plus the associated inverse Compton emission produced by a compact jet, with a marginal or negligible thermal contribution from a cold and/or truncated accretion disc. A theoretical approach based on a jet plus disc model in \citet{markoff2005} has been applied in detail to the cases of M87 and NGC\,1052 \citep{prieto2016,reb2018}, which has served to confirm their nature as pure jet emission. In this work, a simpler approach has been followed by fitting a combination of power laws to describe the jet non-thermal emission. Still, the step spectral slopes found in the optically thin part of the synchrotron spectrum ($-3.7 \lesssim \alpha_0 \lesssim -1.3$ in the IR to optical/UV) are not easily explained by conventional jet models assuming a canonical energy distribution for the accelerated particles. For instance, $N(E) \propto E^{-p}$ with $p \sim 2.4$ leads to an expected $\alpha_0 = (1 - p)/2 \sim -0.7$. In contrast, the spectral slopes observed in our LLAGN sample would correspond to $3.6 \lesssim p \lesssim 8.4$, requiring an extremely efficient --\,and possibly unphysical\,-- cooling mechanism. Alternatively, steep $\alpha_0$ values could be explained if the particles in the jet-corona have not been efficiently accelerated, that is, if the particles follow an energy distribution close to a relativistic maxwellian, instead of a power law, known as the Maxwell-J\"uttner distribution \citep[e.g.][]{jones1979,falcke2000,shahbaz2013}. These `thermalised electrons', located at the jet base or corona, would contribute significantly or even dominate the optically thin synchrotron continuum in the IR to optical/UV range, as shown by \citet{falcke2000} for the case of Sgr A$^*$ and by \citet{plotkin2015} for the case of quiescent black hole X-ray binaries.

\begin{figure}
\centering
\includegraphics[width = \columnwidth]{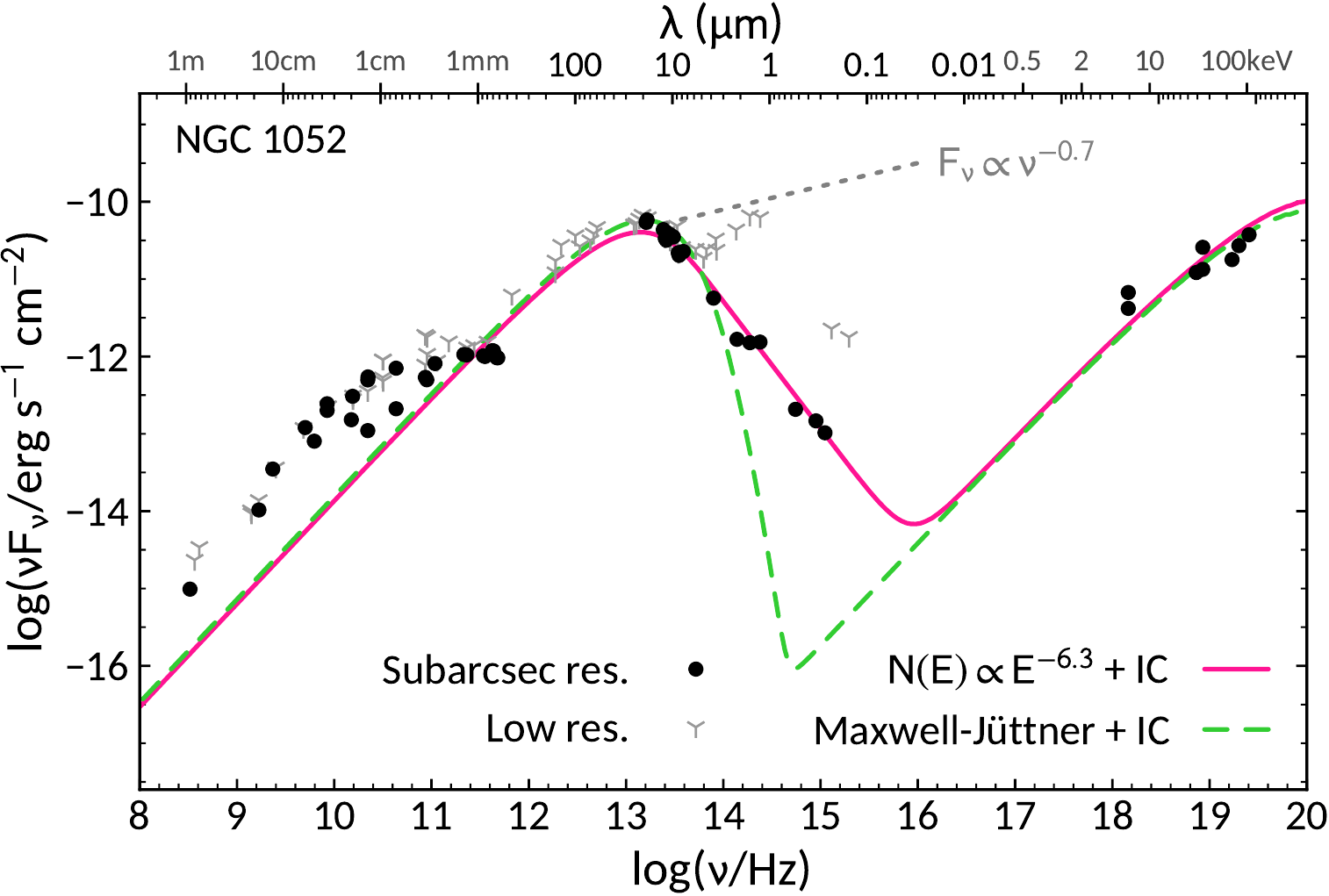}
\caption{Observed SED for NGC\,1052 at subarsec (black circles) and low-angular resolution (grey spikes) compared with the predicted self-absorbed synchrotron plus associated inverse Compton (IC) emission for two particle distributions: a Maxwell-J\"uttner energy distribution (dashed green line), representative of a thermalised distribution of relativistic electrons; and a power law energy distribution with a steep index, $p = 6.3$ (solid pink line; see Eq.\,\ref{eq_E}). For comparison, the continuum spectrum expected for a power-law energy distribution with $p = 2.4$ ($F_\nu \propto \nu^{-0.7}$) is also shown (dotted grey line).}\label{fig_IC}
\end{figure}

To investigate the origin of the optically thin emission, we compare the SED of NGC\,1052 with the expected spectrum radiated by thermalised electrons, which is described by a Maxwell-J\"uttner distribution \citep{zaninetti2020}. An electron temperature of $T_{\rm e} = 10^{12}\, \rm{K}$ and an isotropic magnetic field with an intensity of $B = 0.7\, \rm{G}$ are adopted to reproduce the synchrotron emission in the radio-to-IR range (Fig.\,\ref{fig_IC}). Then, it is necessary to calculate the seed photon density field in order to compute the inverse Compton component. For this purpose, we assume a uniform spherical source with radius $r = 1700\, \rm{AU}$, adopting as scattering particles the same electron distribution that produces the synchrotron radiation. The magnetic field intensity and the source radius considered here are similar to those measured in LLAGN, for instance, $0.26\, \rm{G}$ for the case of Centaurus\,A in \citet{meisenheimer2007}, and $\sim 700\, \rm{AU}$ for the diameter of the central compact radio source in M87 \citep{eht2019a}. The associated inverse Compton emission (synchrotron self-Compton in this case) is then computed taking into account the monochromatic differential cross section of \citet{aharonian1981} and the analytical approximations by \citet{2014ApJ...783..100K}, implemented in the python package \textsc{naima}\footnote{\url{https://naima.readthedocs.io/en/latest/index.html}} \citep{2015ICRC...34..922Z}.

Fig.\,\ref{fig_IC} shows the sub-arcsecond and low-angular resolution SED of NGC\,1052, compared with the model predictions. The thermal electron distribution (dashed green line) is able to reproduce the submm to mid-IR continuum and the X-ray continuum. At lower frequencies the optical depth effects and the contribution from the inner extended jet can introduce large deviations, but the model prediction is consistent with the data when the unresolved radio core dominates the continuum emission. However, the model significantly underestimates the observed SED in the near-IR to UV range, due to the sharp cutoff of the thermal distribution shortwards of the peak power emission at $\sim 20\, \rm{\micron}$. Thus, an additional contribution to the high energy tail of the particle distribution is still required to reproduce the optically thin component \citep[e.g.][]{wardzinski2001,veledina2011}. On the other hand, the X-ray continuum ($0.1$--$3 \times 10^{19}\, \rm{Hz}$) is well reproduced by the thermalised electron distribution. X-ray photons for this model are originated by upscattered seed photons with frequencies of about $0.3$--$8 \times 10^{13}\, \rm{GHz}$ ($100$--$3.4\, \rm{\micron}$), given the average Lorentz factor $\gamma_e = 506$ value for the electron distribution. Thus, the lack of UV photons predicted for this model would only affect the predicted inverse Compton continuum above $\gtrsim 10^{20}\, \rm{Hz}$. For a more detailed study of the physics of the thermal synchrotron radiation we refer to \citet{wardzinski2000}.

To reproduce also the IR-to-optical/UV continuum, we adopted a different energy distribution for the electrons, consisting on a power-law component plus an exponential cutoff:
\begin{equation}\label{eq_E}
  f(E) = A \left(\frac{E}{E_0} \right)^{-p} e^{-\left(\rfrac{E}{E_{\rm cut}} \right)^\beta}, %\exp(-\left(\frac{E}{E_{\rm cut}} \right)^\beta)
\end{equation}
where $A$ is a normalisation constant, $E_0 = 10\, \rm{GeV}$, $p = 6.3$, $E_{\rm cut} = 100\, \rm{TeV}$, and $\beta = 2$. In this case the magnetic field intensity adopted is slightly lower, $B = 0.5\, \rm{G}$, keeping the same core radius of $r = 1700\, \rm{AU}$ to derive the inverse Compton component. The high $p$ value adopted provides an extended tail of electrons with high energies when compared to the Maxwell-J\"uttner distribution, allowing us to reproduce the IR to optical/UV continuum (solid pink line in Fig.\,\ref{fig_IC}). As a consequence, the magnetic field intensity required to reproduce the X-ray emission is slightly lower. The fraction of energy radiated by the non-thermal tail of this distribution is 21\% of the total luminosity in the case of NGC\,1052.

This analysis suggests that the particles producing the compact jet emission in LLAGN are mostly thermalised, following a relativistic Maxwell distribution centred at a considerably high temperature to explain the high turnover frequencies. Nevertheless, some extended high-energy tail is still required to explain the near-IR to optical/UV continuum. A similar scenario has been proposed for the case of compact jets in black hole X-ray binaries in quiescence \citep{shahbaz2013,plotkin2015}. This is in contrast with the characteristic power-law index for efficiently accelerated particles in bright radio galaxies ($p \sim 2.4$), which leads to a much flatter optically thin synchrotron continuum ($\alpha_0 \sim -0.7$; dotted-grey line in Fig.\,\ref{fig_IC}). The presence of an extended non-thermal tail in the electron energy distribution has also been suggested in LLAGN based on the relation between the Eddington ratio and the slope of the X-ray spectrum \citep{niedzwiecki2014,niedzwiecki2015}. These non-thermal electrons could be originated by, for instance, pion decays in proton--proton interactions \citep{mahadevan1999,niedzwiecki2015}, or alternatively by direct acceleration of the electrons due to energy dissipation caused by magnetic field reconnection \citep{riquelme2012}.

The continuum spectral shapes of LLAGN in our sample share some remarkable similarities with the core of peaked-spectrum radio sources. These are compact and powerful radio sources in the nuclei of galaxies with a continuum spectrum characteristic of self-absorbed synchrotron emission \citep[see][and references therein]{odea1998,odea2021}. That is, showing a flat spectrum at low radio frequencies and a spectral turnover typically in the MHz to GHz range, followed by an optically thin steep continuum a higher frequencies. The population of high-frequency peakers --\,those sources with a spectral turnover above $\gtrsim 10\, \rm{GHz}$\,-- was discovered last due to lack of systematic surveys in this spectral range \citep{hancock2010,bonaldi2013}. Overall, peaked-spectrum sources are considered as young radio active AGN, due to the inverse relation between the turnover frequency and the size of the radio source \citep{odea1997,orienti2016}. Sources with the youngest spectral ages usually show the most compact sizes ($< 500\, \rm{pc}$) and the highest turnover frequencies ($10\, \rm{GHz}$; \citealt{tinti2005}), as has been confirmed by VLBI observations \citep{hancock2009}. In this context, the spectral age refers to the lifetime of the synchrotron emitting electrons, which can be associated with the last episode of jet activity in these nuclei \citep{murgia2003,odea1998,orienti2007}. The large number density of compact young radio sources suggests that the majority of them do not evolve into extended radio galaxies \citep{kunert2010}. Additionally, SDSS spectroscopy of a sample of peaked-spectrum sources shows that a large fraction of them ($\sim 1/3$) reside in LLAGN \citep{liao2020}. Nevertheless, peaked-spectrum radio sources are also frequently found among quasars and narrow-line Seyfert 1 galaxies \citep{wu2009,singh2018}, which suggests that they are a heterogeneous class and are not particularly associated with a certain type of active nucleus.

From this perspective, the cores of the LLAGN in our sample can be considered as extreme frequency peakers, since their spectral turnover is found at $\nu_{\rm b} > 210\, \rm{GHz}$ (see Table\,\ref{tab_fit}) and their sizes are not resolved by \textit{HST} or VLT/NaCo ($\lesssim 10\, \rm{pc}$), in some cases not even by VLTI/MIDI interferometry ($< 0.5\, \rm{pc}$ in NGC\,1052, \citealt{jafo2019}). Except for M87, NGC\,4261 and NGC\,1052, which show extended radio structures, the rest of LLAGN in our sample are compact radio sources \citep{mezcua2014}. As a matter of fact, NGC\,1052 is considered as the closest GHz-peaked source \citep[e.g.][]{tingay2015} due to the peak emission seen around $\sim 10\, \rm{GHz}$ in Fig.\,\ref{fig_seds}, although the radio core is only detected at much higher frequencies \citep{baczko2016,pasetto2019} and peaks in the mid-IR (Fig.\, \ref{fig_seds}).

The LLAGN in our sample are also comparable to FR\,0 sources. These are compact radio sources similar to the core of FR\,I galaxies but with much fainter extended emission in the adjacent radio lobes, which can be completely missing \citep{baldi2015}. The majority of FR\,0 nuclei show low excitation ratios in their optical emission lines and belong to the LINER class \citep{baldi2018}, with most of them showing spectral turnovers at high radio frequencies \citep{sadler2014,tingay2015}, although their luminosities are significantly lower when compared to the core of GHz-peaked sources.

Overall, LLAGN show a similar continuum shape but higher turnover frequencies when compared to young radio sources, while the steep optically thin power-law continuum in LLAGN suggests a dearth of accelerated particles in these nuclei. These characteristics may be indicative of the presence of undeveloped jets in LLAGN, that is the manifestation of quiescent jet cores whose continuum emission is dominated by self-Compton quasi-thermal synchrotron radiation across the whole electromagnetic spectrum. These characteristics can be associated with a quiescent state that may precede the development of a new jet event. This scenario is arguably similar to the case of black hole X-ray binaries, which show a much weaker jet particle acceleration in the quiescence state that follows a brighter more active phase \citep{plotkin2015}. Then, internal shocks and magnetic reconnection events can lead to efficient particle acceleration and thus launch outflows of relativistic plasma \citep[e.g.][]{marscher1985,sironi2009,sironi2015,blandford2019}, that allow the formation of strongly collimated jets \citep[e.g.][]{boccardi2016,boccardi2021}. The large number density of LLAGN in the local Universe when compared to that of radio galaxies suggests that most of these quiescent jets do not develop bright extended radio structures in the innermost parsecs. This is in agreement with the trend found by \citet{lister2019} for the MOJAVE AGN survey, in which sources with the highest synchrotron turnover frequencies also show the slowest jet propagation speeds. Following similar arguments as in the case of the core of peaked-spectrum radio sources \citep[see][]{odea2021}, this could be explained by the presence of dense gas in the nuclear environment that confines the jet and prevents its expansion \citep[e.g. the jet-ISM interaction in NGC\,1386][]{rodriguez-ardila2017}, and/or by intermittent activity in short-lived events unable to develop extended radio structures. The former scenario might not apply to sources in which previous jet activity have already carved a channel through the ISM (e.g. NGC\,4261), nevertheless changes in the orientation between different jet episodes may keep the restarted jet confined within the innermost few parsecs \citep[e.g. NGC\,1052;][]{kadler2004}.

Nevertheless, some of the nuclei in our sample show extended radio emission over kiloparsec scales, indicative of stronger activity in the past (e.g. NGC\,1052, M87, and NGC\,4261). Jets are known to have an episodic behaviour \citep{baum1990,stanghellini2005,shabala2008,an2012,shulevski2012,singh2016,tadhunter2016}, with cyclic activity events each lasting a few tens of Myr, as suggested by the multiple generations of X-ray cavities surrounding radio-loud nuclei \citep{mcnamara2007,vantyghem2014} and the fraction of restarted radio galaxies with respect to the remnant or inactive population \citep{jurlin2020}. Thus, we interpret the presence of compact jet cores in LLAGN as a low state of nuclear activity that may happen before or amidst major radio-loud events.

An important implication of this scenario is that most of the energy in LLAGN may be released through the jet kinetic power, while radiative processes are modest in comparison. This can be substantially relevant for galaxy evolution studies, since low-power radio jets may represent a major but underestimated source of kinetic feedback in galaxies \citep[e.g.][]{may2018,tabatabaei2018,jafo2020,shi2022,papachristou2021,girdhar2022}.

\section{Summary}\label{summary}

Low-luminosity active galactic nuclei (LLAGN) are key sources among the AGN family because they allow us to study accretion physics in supermassive black holes in the sub-Eddington domain ($L \lesssim 10^{-3}\, \rm{L_{Edd}}$), when the accretion process becomes radiatively inefficient due to profound changes affecting the accretion disc properties in this regime. We have analysed a sample of nine prototypical LLAGN in the nearby Universe ($3$--$32\,\rm{Mpc}$), accessible with adaptive optics observations in the near-IR that allow us to probe the complete mid-IR to optical/UV range and build a multi-wavelength spectral energy distribution (SED) at sub-arcsecond resolution, including radio, millimetre/submillimetre and X-ray measurements from the literature. The high-angular resolution in the IR-to-optical/UV range is critical to isolate the nuclear emission and avoid the starlight contamination from the host galaxy. We identify the IR nucleus in 8 out of 9 targets, the exception is the intermediate-mass black hole in NGC\,404, whose IR and optical continuum seems to be significantly fainter than the stellar emission in this range.

The continuum spectra of the central few parsecs of the LLAGN sample in this work are found to present a remarkable similar spectrum, featureless in shape over 10 orders of magnitude in frequency. The nuclear continua for the 8 LLAGN detected in the IR and optical range show the distinctive spectral shape of self-absorbed synchrotron emission in compact jets. This is characterised by a relatively flat spectrum at radio frequencies, where the emission is optically thick, followed by a turnover at millimetre/IR wavelengths and a steep continuum with $-3.7 < \alpha_0 < -1.3$ ($F_\nu \propto \nu^{\alpha_0}$) in the optically thin domain that extends (at least) to the far-UV range for non-obscured sources. This scenario is further supported by the polarisation properties in the cases of M87 and NGC\,1052. Based on the analysis of the emission line ratios and photo-ionisation simulations, we conclude that the extreme UV continuum is dominated by the inverse Compton emission, which rises in this spectral range up to the X-ray range and possibly higher energies. This implies that the nebular gas in the vicinity of the LLAGN in our sample ($\lesssim 10\, \rm{pc}$) is mainly ionised by synchrotron emission radiated by a compact jet, specifically by the upscattered synchrotron photons produced by inverse Compton processes in the black hole corona and/or the jet base. Nevertheless, shocks may still have an important contribution at larger distances, especially in the presence of outflows. Evidence of the accretion disc emission is found only for the Sombrero galaxy (NGC\,4594), although the thermal-like bump detected in this case peaks at $\sim 1\, \rm{\micron}$, suggesting the presence of a very cold and/or a truncated accretion disc that has a negligible impact on the structure of the ionised gas.

The compact jet continuum shape in LLAGN is in agreement with both the $\alpha_{\rm ox}$ spectral index in the optical to X-ray range and the mid-IR to X-ray correlation observed in the brighter Seyfert nuclei. The generalisation of the origin of the IR to X-ray correlation in AGN as caused by dust reprocessed emission from the accretion disc is not correct. LLAGN fall exactly on the same correlation, and yet, the cause for it is the Comptonisation of the jet emission in the IR upscattered into the X-rays. This result shows that these correlations are not necessarily associated with a single single physical process, but different mechanisms involving energy balance between the IR or the optical and the X-rays may drive the observed fluxes in these ranges into the same correlation.

A distinctive characteristic of LLAGN is their steep optically thin continuum ($-3.7 < \alpha_0 < -1.3$), in contrast with other jet-dominated sources such as radio galaxies ($\alpha_0 \sim -0.7$). This can be explained by synchrotron emission from electrons at the jet base or corona that follow a relativistic quasi-thermal energy distribution with a considerably high temperature to explain the high turnover frequencies observed, whereas only $\sim 20$\% of the energy would be radiated by particles in a high-energy power-law tail in the near-IR to optical/UV continuum. These accelerated particles could be produced by pion decays in proton-proton interactions or alternatively by the energy dissipation caused by magnetic field reconnection events in the vicinity of the black hole.

From this study we conclude that the continuum emission of LLAGN is dominated across the whole electromagnetic spectrum by the emission from compact jet cores, which could be regarded as undeveloped or quiescent jets due to the lack of a significant fraction of accelerated particles at their jet base. This implies that most of the energy in these sources is released by the jet kinetic power, suggesting that feedback from LLAGN may be largely underestimated in the context of galaxy evolution. The injection of energy through shocks in these nuclei could re-accelerate the electron population and trigger further jet activity in the future. However, to produce extended jet emission, for instance, up to UV wavelengths as in the case of M87, would require both a sustained period of activity and a relatively low density environment that permits the jet expansion. Extended radio emission over kiloparsec scales is detected for some of these objects (e.g. NGC\,1052, M87, and NGC\,4261), suggesting that stronger jet activity happened in the recent past of these nuclei.

In a short timescale the \textit{James Webb Space Telescope} (\textit{JWST}) will open a new window in the study of LLAGN, through the characterisation of their nuclear IR continua at small angular scales and the spectral diagnostics based on the emission lines from the nebular gas regions around these nuclei. This is one of the main scientific goals of the ReveaLLAGN project \citep{seth2021}, which has been awarded with 33 hours in the \textit{JWST} GO cycle 1 to obtain the nuclear spectra of seven prototypical LLAGN in the $1.7$--$28\, \rm{\micron}$ range.

%% The spectral slope of this LLAGN sample in the X-rays as inferred from the modelling of the SED is ~ -0.7 with a spread of about 0.2. LLAGN constitutes 2/3 of the AGN population in the near Universe, the contribution of this population to the CXB at low luminosities, should be reviewed on the basis of this flatter spectrum as compared with the usual AGN model template with photon index -1.9 (Ueda+03; Gilli+06).

%%%%%%%%%%%%%%%%%%%%%%%%%%%%%%%%%%%%%%%%%%%%%%%%%%%%%%%%%%%%%%
\begin{acknowledgements}
%The authors acknowledge the referee for his/her useful comments that helped to improve the manuscript.
The authors would like to thank the anonymous referee for his/her useful comments and suggestions. The authors acknowledge J.C. Guerra for providing the reduced LBT/PISCES adaptive optics images for NGC\,404. The authors would also like to thank S. Bianchi for his insightful comments on the manuscript. Part of this work was carried out by XLL for his master thesis, which was developed within the PARSEC project at IAC and presented at the ULL. JAFO acknowledges the financial support from the Spanish Ministry of Science and Innovation and the European Union -- NextGenerationEU through the Recovery and Resilience Facility project ICTS-MRR-2021-03-CEFCA. AP thanks the support of the Excellence Cluster ORIGINS which is funded by the Deutsche Forschungsgemeinschaft (DFG, German Research Foundation) under Germany Excellence Strategy EXC-2094-390783311. This work is based in part on observations made with the \textit{Spitzer Space Telescope}, obtained from the NASA/IPAC Infrared Science Archive, both of which are operated by the Jet Propulsion Laboratory, California Institute of Technology under a contract with the National Aeronautics and Space Administration.
\end{acknowledgements}

%%%%%%%%%%%%%%%%%%%%%%%%%%%%%%%%%%%%%%%%%%%%%%%%%%%%%%%%%%%%%%
\bibliographystyle{aa}
\bibliography{llagn}
%%%%%%%%%%%%%%%%%%%%%%%%%%%%%%%%%%%%%%%%%%%%%%%%%%%%%%%%%%%%%%

%\onecolumn
%\appendix
%\label{lastpage}

%In Tables\,\ref{sed_high} \& \ref{sed_low} we show the spectral flux distribution for the nucleus of NGC\,1052 compiled originally in ... Tabulated values have not been corrected for Galactic reddening.

%\begin{table}

%\section{The spectral flux distribution of NGC 1052}\label{app_sed}
%   \tiny
%   \centering
%   \setlength{\tabcolsep}{3.pt}
%   \caption{Spectral flux distribution for the nucleus of NGC\,1052 at sub-arcsecond aperture resolution flux. Tabulated values have not been corrected for Galactic reddening. Intrinsic X-ray fluxes are derived from apertures of a few to several arcseconds in size, however the host galaxy contribution in this range is negligible, thus the fluxes are assumed as representative from the nuclear source. The asterisk symbol in the flux column denotes the presence of multiple jet components resolved within the inner $0.\!\!^{\prime\prime}15$ in the $2.3$--$89\, \rm{GHz}$ range. In this case the tabulated fluxes correspond to the sum of all the jet components.}\label{sed_high}
%   \begin{tabular}{rcccll}
%     \input{tabs/n1052_high.tex}
%   \end{tabular}
%\end{table}

%\clearpage

%\begin{table}
%   \tiny
%   \centering
%   \setlength{\tabcolsep}{3.pt}
%   \caption{Spectral flux distribution for the nucleus of NGC\, 1052 extracted from large aperture ($\gtrsim 1''$) flux measurements.}\label{sed_low}
%   \begin{tabular}{rccll}
%     \input{tabs/n1052_low.tex}
%   \end{tabular}
%\end{table}

\end{document}